



\input epsf.tex
\input harvmac.tex




\def\IC{\relax{\rm I\kern-.18em C}}

\def\IL{\relax{\rm I\kern-.18em L}}
\def\IH{\relax{\rm I\kern-.18em H}}
\def\IR{\relax{\rm I\kern-.18em R}}
\def\IC{\relax\hbox{$\inbar\kern-.3em{\rm C}$}}
\def\IZ{\relax\ifmmode\mathchoice
{\hbox{\cmss Z\kern-.4em Z}}{\hbox{\cmss Z\kern-.4em Z}}
{\lower.9pt\hbox{\cmsss Z\kern-.4em Z}}
{\lower1.2pt\hbox{\cmsss Z\kern-.4em Z}}\else{\cmss Z\kern-.4em Z}\fi}
\def\CA{{\cal A}}

\def\CM {{\cal M}}
\def\CN {{\cal N}}
\def\CR {{\cal R}}
\def\CD {{\cal D}}

\def\CP {{\cal P }}
\def\CL {{\cal L}}

\def\CE {{\cal E}}
\def\CG {{\cal G}}
\def\CH {{\cal H}}

\font\manual=manfnt \def\dbend{\lower3.5pt\hbox{\manual\char127}}

\def\c{\check}
\def\IZ{\relax\ifmmode\mathchoice
{\hbox{\cmss Z\kern-.4em Z}}{\hbox{\cmss Z\kern-.4em Z}}
{\lower.9pt\hbox{\cmsss Z\kern-.4em Z}}
{\lower1.2pt\hbox{\cmsss Z\kern-.4em Z}}\else{\cmss Z\kern-.4em Z}\fi}
\def\half {{1\over 2}}
\def\sdtimes{\mathbin{\hbox{\hskip2pt\vrule height 4.1pt depth -.3pt width
.25pt
\hskip-2pt$\times$}}}
\def\p{\partial}

\def\CL {{\cal L}}

\def\CM {{\cal M}}
\def\CN {{\cal N}}

\def\CP {{\cal P }}
\def\CQ {{\cal Q }}
\def\CE{{\cal E }}

\def\CU{{\cal U}}

\def\Det{{\rm Det}}


\def\IZ{\relax\ifmmode\mathchoice
{\hbox{\cmss Z\kern-.4em Z}}{\hbox{\cmss Z\kern-.4em Z}}
{\lower.9pt\hbox{\cmsss Z\kern-.4em Z}}
{\lower1.2pt\hbox{\cmsss Z\kern-.4em Z}}\else{\cmss Z\kern-.4em
Z}\fi}
\def\IB{\relax{\rm I\kern-.18em B}}
\def\IC{{\relax\hbox{$\inbar\kern-.3em{\rm C}$}}}
\def\ID{\relax{\rm I\kern-.18em D}}
\def\IE{\relax{\rm I\kern-.18em E}}
\def\IF{\relax{\rm I\kern-.18em F}}
\def\IG{\relax\hbox{$\inbar\kern-.3em{\rm G}$}}
\def\IGa{\relax\hbox{${\rm I}\kern-.18em\Gamma$}}
\def\IH{\relax{\rm I\kern-.18em H}}
\def\II{\relax{\rm I\kern-.18em I}}
\def\IK{\relax{\rm I\kern-.18em K}}
\def\IP{\relax{\rm I\kern-.18em P}}

\def\IQ{\relax\hbox{$\inbar\kern-.3em{\rm Q}$}}
\def\IP{\relax{\rm I\kern-.18em P}}

\def\IB{\relax{\rm I\kern-.18em B}}
\def\ID{\relax{\rm I\kern-.18em D}}
\def\IE{\relax{\rm I\kern-.18em E}}
\def\IF{\relax{\rm I\kern-.18em F}}
\def\IG{\relax\hbox{$\inbar\kern-.3em{\rm G}$}}
\def\IGa{\relax\hbox{${\rm I}\kern-.18em\Gamma$}}
\def\IH{\relax{\rm I\kern-.18em H}}
\def\II{\relax{\rm I\kern-.18em I}}
\def\IJ{\relax{\rm I\kern-.18em J}}
\def\IK{\relax{\rm I\kern-.18em K}}
\def\IL{\relax{\rm I\kern-.18em L}}

\def\IN{\relax{\rm I\kern-.18em N}}
\def\IO{\relax{\rm I\kern-.18em O}}
\def\IP{\relax{\rm I\kern-.18em P}}
\def\IQ{\relax\hbox{$\inbar\kern-.3em{\rm Q}$}}
\def\IR{\relax{\rm I\kern-.18em R}}
\def\IW{\relax\hbox{$\inbar\kern-.3em{\rm W}$}}

\def\im{{\rm Im}}

\def\inbar{\,\vrule height1.5ex width.4pt depth0pt}

\def\ndt{\noindent}
\def\p{\partial}

\font\cmss=cmss10 \font\cmsss=cmss10 at 7pt
\def\IR{\relax{\rm I\kern-.18em R}}

\def\sdtimes{\mathbin{\hbox{\hskip2pt\vrule
height 4.1pt depth -.3pt width .25pt\hskip-2pt$\times$}}}
\def\Tr{\rm Tr} 

\def\vol{{\rm vol}}

\def\bgn{\bigskip\ndt}

\def\bul{$\bullet$}

\def\inv{^{\raise.15ex\hbox{${\scriptscriptstyle -}$}\kern-.05em 1}}

\def\Dsl{\,\raise.15ex\hbox{/}\mkern-13.5mu D} 
\def\dsl{\raise.15ex\hbox{/}\kern-.57em\partial}

\def\l{\ell}

\def\tr{{\rm tr}} \def\Tr{{\rm Tr}}


\def\E{{\bf E}}

\def\R{{\bf R}}

\def\Z{{{\bf Z}}}

\def\half{{{1\over 2}}}

\def\CE{{\cal E}}
\def\CP{{\cal P}}

\def\a{{\alpha}}

\def\Z{{\bf Z}}

\def\CM{{{\cal M}}}

\def\CR{{\cal R}}

\def\CN{{{\cal N}}}

\def\p{{\partial}}



\def\boxit#1{\vbox{\hrule\hbox{\vrule\kern8pt
\vbox{\hbox{\kern8pt}\hbox{\vbox{#1}}\hbox{\kern8pt}}
\kern8pt\vrule}\hrule}}
\def\mathboxit#1{\vbox{\hrule\hbox{\vrule\kern8pt\vbox{\kern8pt
\hbox{$\displaystyle #1$}\kern8pt}\kern8pt\vrule}\hrule}}


\lref\booth{P. Booth, P. Heath, C. Morgan, and R. Piccinini, 
``Remarks on the homotopy type of groups of gauge transformations,'' 
C.R. Math. Rep. Acad. Sci. Canada {\bf 3}(1981)3} 
\lref\sutherland{W.A. Sutherland, ``Function spaces related to 
gauge groups,'' Proc. Roy. Soc. Edinburgh, {\bf 121A}(1992)185}

\lref\DaiKQ{
X.~Dai and D.~S.~Freed,
``Eta-invariants and determinant lines,''
J.\ Math.\ Phys.\  {\bf 35}, 5155 (1994)
[Erratum-ibid.\  {\bf 42}, 2343 (2001)]
[arXiv:hep-th/9405012].
}

\lref\DiaconescuWY{
D.~E.~Diaconescu, G.~W.~Moore and E.~Witten,
``E(8) gauge theory, and a derivation of K-theory from M-theory,''
arXiv:hep-th/0005090.
}

\lref\DiaconescuWZ{
D.~E.~Diaconescu, G.~W.~Moore and E.~Witten,
``A derivation of K-theory from M-theory,''
arXiv:hep-th/0005091.
}

\lref\DFM{D.E. Diaconescu, D. Freed, and G. Moore, to appear, we hope.} 
\lref\DMW{D.E.Diaconescu, G.Moore, E. Witten,
''E8 Gauge Theory, and a Derivation of K-Theory from M-Theory'',
hep-th/005090 }
\lref\DiaconescuWY{
D.~E.~Diaconescu, G.~W.~Moore and E.~Witten,
``E(8) gauge theory, and a derivation of K-theory from M-theory,''
Adv.\ Theor.\ Math.\ Phys.\  {\bf 6}, 1031 (2003)
[arXiv:hep-th/0005090].
}

\lref\DMWii{D.E. Diaconescu, G. Moore, and E. Witten, 
``A Derivation of K-Theory from M-Theory,'' hep-th/0005091}

\lref\DiaconescuWZ{
D.~E.~Diaconescu, G.~W.~Moore and E.~Witten,
``A derivation of K-theory from M-theory,''
arXiv:hep-th/0005091.
}

\lref\DMWUN{D.E.Diaconescu, G.Moore, E. Witten, unpublished.}
\lref\DM{D.E. Diaconescu and G. Moore, unpublished. } 
\lref\MW{ G.Moore, E. Witten, ``Integration over the u-plane in
Donaldson theory'',hep-th/9709193}

\lref\FreedTG{
D.~Freed, J.~A.~Harvey, R.~Minasian and G.~W.~Moore,
``Gravitational anomaly cancellation for M-theory fivebranes,''
Adv.\ Theor.\ Math.\ Phys.\  {\bf 2}, 601 (1998)
[arXiv:hep-th/9803205].
}
\lref\MooreNN{
G.~W.~Moore,
``Some comments on branes, G-flux, and K-theory,''
Int.\ J.\ Mod.\ Phys.\ A {\bf 16}, 936 (2001)
[arXiv:hep-th/0012007].
}

\lref\freed{D. Freed,
Dirac Charge Quantization and Generalized Differential Cohomology,''
hep-th/0011220 } 
\lref\Hull{C.M. Hull,''Massive String Theories From M-Theory and F-Theory'',
JHEP 9811 (1998) 027,hep-th/
9811021 }
\lref\Town{C.Hull,P.Townsend,''Unity of Superstring Dualities,''
Nucl.Phys. B438 (1995) 109-137,hep-th/9410167}
\lref\GPR{ A. Giveon, M. Porrati, E. Rabinovici,
'' Target Space Duality in String Theory'', hep-th/9401139 }
\lref\BRG{E.Bergshoeff,M.de Roo,M.B.Green, G.Papadopoulos,
P.K.Townsend,''Duality of II 7-branes and 8-branes'', hep-th/9601150}
\lref\Lavr{I.Lavrinenko, H.Lu, C.Pope,
T.Tran,``U-duality as general Coordinate Transformations, and Spacetime
Geometry'', hep-th/9807006 }
\lref\Cvet{ M. Cvetic, H. Lu, C.N. Pope, K.S. Stelle,''
T-Duality in the Green-Schwarz Formalism, 
and the Massless/Massive IIA Duality Map'',Nucl.Phys. B573 (2000) 149-176,
 hep-th/9907202}
\lref\Piol{E.Kiritsis and B. Pioline, ``On $R^4$ threshhold corrections
in IIB string theory and (p,q) string instantons,'' 
Nucl. Phys. {\bf B508}(1997)509; 
B. Pioline, H. Nicolai, J. Plefka, A. Waldron,
 $R^4$ couplings, the fundamental membrane and 
exceptional theta correspondences, hep-th/0102123 }  
\lref\mm{R. Minasian and G. Moore,``K Theory and Ramond-Ramond Charge,''
JHEP {\bf 9711}:002, 1997; hep-th/9710230.}

\lref\mw{G. Moore and E. Witten, ``Self-Duality, Ramond-Ramond Fields, and K-Theory,''
hep-th/9912279;JHEP 0005 (2000) 032}
\lref\Warn{C.Isham,C.Pope,N.Warner,
''Nowhere-vanishing spinors and triality rotations in 8-manifolds,''
Class.Quantum Grav.5(1988)1297}
%

%
\lref\WittenMD{
E.~Witten,
``On flux quantization in M-theory and the effective action,''
J.\ Geom.\ Phys.\  {\bf 22}, 1 (1997)
[arXiv:hep-th/9609122].
}
\lref\WittenBT{
E.~Witten,
``Topological Tools In Ten-Dimensional Physics,''
Int.\ J.\ Mod.\ Phys.\ A {\bf 1}, 39 (1986).
}
\lref\WittenHC{
E.~Witten,
``Five-brane effective action in M-theory,''
J.\ Geom.\ Phys.\  {\bf 22}, 103 (1997)
[arXiv:hep-th/9610234].
}
\lref\WittenVG{
E.~Witten,
``Duality relations among topological effects in string theory,''
JHEP {\bf 0005}, 031 (2000)
[arXiv:hep-th/9912086].
}

\lref\wittenk{E. Witten, ``$D$-Branes And $K$-Theory,''
JHEP {\bf 9812}:019, 1998; hep-th/9810188.}      
\lref\wittenvbr{E.Witten,'' Five-brane effective action in M-theory,''
J.Geom.Phys.,22 (1997)103,hep-th/9610234}

\lref\wittenduality{E. Witten, ``Duality Relations Among Topological Effects In String Theory,''
hep-th/9912086;JHEP 0005 (2000) 031} 
\lref\wittenstrings{E. Witten, ``Overview of K-theory applied to
strings,'' hep-th/0007175.}          
\lref\Sethi{ S. Sethi, C. Vafa, E. Witten,
``Constraints on Low-Dimensional String Compactifications'',
hep-th/9606122, Nucl.Phys. B480 (1996) 213-224 }
\lref\Wits{E.Witten,''On S-Duality in Abelian Gauge Theory'',
hep-th/9505186}
\lref\Verl{E. Verlinde,''Global Aspects of Electric-Magnetic Duality'',
Nucl.Phys. B455 (1995) 211-228,hep-th/9506011}
\lref\Romans{L. Romans,''Massive IIA Supergravity in Ten Dimensions',
Phys.Let.169B(1986)374}
\lref\Myers{P. Meessen, T. Ortin,
''An Sl(2,Z) Multiplet of Nine-Dimensional Type II Supergravity Theories'',
Nucl.Phys. B541 (1999) 195-245, hep-th/9806120 }
\lref\Siegel{W.Siegel,''Hidden ghosts'', Phys.Lett.B93(1980),170 }
\lref\Deser{S.Deser,P.van Nieuwenhuizen,''One-loop
divergences of quantized Einstein-maxwell fields'',Phys.Rev.D(1974),
v.10,p.401}
\lref\Hooft{G.'t Hooft, M.Veltman,''One loop divergences in
the theory of gravitation'', Annales 
Inst.Henri Poincare ,vol.20,(1974),69}
\lref\CJS{E.Cremmer, B.Julia, J.Scherk, 
``Supergravity theory in 11 dimensions'',
Phys.Lett. B76(1978)409}
\lref\Nielsen{N.Nielsen, 
 ``Ghost counting  in supergravity'', NPB 140(1978)499}
\lref\Kallosh{R.Kallosh,''Modified  Feynman
rules in supergravity'',NPB 141(1978)141}
\lref\Fuj{K.Fujikawa,''
Path  Integral for gauge theories with fermions'',
Phys. Rev.D.21(1980) 2848}
\lref\Ibusa{J.Igusa, {\it Theta-functions}, Berlin, New York, 
Springer-Verlag, 1972.}
\lref\Phong{E.D'Hoker,D.Phong,''Conformal scalar fields 
and chiral splitting on SuperRiemann
surfaces'',Comm.Math.Phys.125
(1989)469;

K.Aoki,E.D'Hoker,D.Phong,
''Unitarity of closed superstring Perturbation Theory'',
NPB 342(1990)149-230 
}
\lref\Geg{J.Gegenborg, G.Kunstatter,''The partition function for
topological field theories``,hep-th/9304016}
\lref\moorestrings{G.Moore,''Some comments on Branes, G-flux, and K-theory'',
Int.J.Mod.Phys.A16   (2001)936-944,
hep-th/0012007} 
\lref\Mathai{V. Mathai, D. Stevenson, 
``Chern character in twisted K-theory: equivariant and holomorphic cases''
hep-th/0201010}
\lref\MathaiB{V. Mathai, R.B. Melrose, and I.M. Singer, 
``The index of projective families of elliptic operators,'' 
math.DG/0206002}
\lref\Rabin{J.Rabin, M.Bergvelt,''Super curves, their Jacobians,
and super KP equations'',alg-geom/9601012}
\lref\Sch{S.Fredenhagen, V.Schomerus, Non-commutative geometry,
gluon condensates, and K-theory'',JHEP 0004(2001)007,
hep-th/0012164}
\lref\Stan{S.Stanciu,''An illustrated guide to D-branes in SU(3)'',
hep-th/0111221}

\lref\MaldacenaXJ{
J.~M.~Maldacena, G.~W.~Moore and N.~Seiberg,
``D-brane instantons and K-theory charges,''
JHEP {\bf 0111}, 062 (2001)
[arXiv:hep-th/0108100].
}

\lref\MMS{J.Maldacena,G.Moore, N.Seiberg, ``D-brane instantons
and K-theory charges'', JHEP 0111(2001)062; hep-th/0108100 }
\lref\mmsiii{J. Maldacena, G. Moore, and N. Seiberg, 
``D-brane charges in five-brane backgrounds,'' JHEP 0110 (2001)
005; hep-th/0108152}

\lref\Mathaii{P.Bouwknegt, V.Mathai,``D-branes, B-fields and 
twisted K-theory'', ~~
JHEP 0003
(2000)007, hep-th/0002023}
\lref\APS{M.Atiyah,V.Patodi, I.Singer,
Math.Proc.Cambridge Phil.Soc.77(1975)43;405.}
\lref\fh{D. S. Freed, M. J. Hopkins, 
``On Ramond-Ramond fields and K-theory,'' JHEP 0005 (2000) 044;
hep-th/0002027}
\lref\birreldavies{N.D. Birrell and P.C.W. Davies, 
{\it Quantum Fields in Curved Space}, Cambridge Univ. Press, 1982}
 
\lref\stong{R. Stong, ``Calculation of
$\Omega_{11}^{spin}(K({\bf Z},4))$'' in
{\it Unified String Theories}, 1985 Santa Barbara
Proceedings, M. Green and D. Gross, eds. World Scientific 1986.}

\lref\asv{M. F. Atiyah, I. M. Singer, ``The index of
elliptic operators: V'' Ann. Math. {\bf 93} (1971) 139.}

\lref\szabo{K. Olsen and R.J. Szabo,
``Constructing $D$-Branes from $K$-Theory,''
hep-th/9907140.  }

\lref\hori{K. Hori, ``D-branes, T-duality, and Index Theory,''
Adv.Theor.Math.Phys. 3 (1999) 281-342; hep-th/9902102  }

\lref\evslin{A.Adams,J.Evslin,''The Loop Group of $E_8$ and 
K-Theory from 11d,''  hep-th/0203218}

\lref\HopkinsRD{
M.~J.~Hopkins and I.~M.~Singer,
``Quadratic functions in geometry, topology, and M-theory,''
arXiv:math.at/0211216.
}

\lref\MooreVF{
G.~Moore,
``K-theory from a physical perspective,''
arXiv:hep-th/0304018.
}

\lref\HoravaMA{
P.~Horava and E.~Witten,
``Eleven-Dimensional Supergravity on a Manifold with Boundary,''
Nucl.\ Phys.\ B {\bf 475}, 94 (1996)
[arXiv:hep-th/9603142].
}

\lref\HoravaQA{
P.~Horava and E.~Witten,
``Heterotic and type I string dynamics from eleven dimensions,''
Nucl.\ Phys.\ B {\bf 460}, 506 (1996)
[arXiv:hep-th/9510209].
}

\lref\FabingerJD{
M.~Fabinger and P.~Horava,
``Casimir effect between world-branes in heterotic M-theory,''
Nucl.\ Phys.\ B {\bf 580}, 243 (2000)
[arXiv:hep-th/0002073].
}

\lref\MeloschWM{
S.~Melosch and H.~Nicolai,
``New canonical variables for d = 11 supergravity,''
Phys.\ Lett.\ B {\bf 416}, 91 (1998)
[arXiv:hep-th/9709227].
}

\lref\lott{J. Lott, ``R/Z index theory,'' Comm. Anal. Geom. 
{\bf 2} (1994) 279} 

\lref\zucchini{R. Zucchini, ``Relative topological integrals and 
relative Cheeger-Simons differential characters,'' hep-th/0010110}

\lref\daifreed{X. Dai and D.S. Freed, ``$\eta$-Invariants and 
Determinant Lines,'' hep-th/9405012; D.S. Freed, ``Determinant 
Line Bundles Revisited,'' dg-ga/9505002}

\lref\horavawitten{Horava and Witten} 

\lref\freeddq{D. Freed, Dirac Quantization ...., hep-th} 

\lref\horava{P. Horava, ``M-Theory as a Holographic Field Theory,'' 
hep-th/9712130} 

\lref\morrison{D. Morrison, Talk at Strings2002, http://www.damtp.cam.ac.uk/strings02/avt/morrison/ }

\lref\sevenauthor{Jan de Boer, Robbert Dijkgraaf, Kentaro Hori, Arjan Keurentjes, 
John Morgan, David R. Morrison, Savdeep Sethi, ``Triples, Fluxes, and Strings,'' hep-th/0103170 ;
Adv.Theor.Math.Phys. 4 (2002) 995-1186}
%
\lref\MooreGB{
G.~W.~Moore and E.~Witten,
``Self-duality, Ramond-Ramond fields, and K-theory,''
JHEP {\bf 0005}, 032 (2000)
[arXiv:hep-th/9912279].
}

\lref\deBoerPX{
J.~de Boer, R.~Dijkgraaf, K.~Hori, A.~Keurentjes, J.~Morgan, D.~R.~Morrison and S.~Sethi,
``Triples, fluxes, and strings,''
Adv.\ Theor.\ Math.\ Phys.\  {\bf 4}, 995 (2002)
[arXiv:hep-th/0103170].
}

\lref\dfmunpub{E. Diaconescu, D. Freed,  and G. Moore, to appear, maybe.} 

\lref\toappear{To appear.}

\lref\freedhopkins{D. Freed and M. Hopkins, to appear} 

\lref\evslin{J. Evslin and U. Varadarajan, 
``K-Theory and S-Duality: Starting Over from Square 3,'' 
hep-th/0112084  ;    Journal-ref: JHEP 0303 (2003) 026}

\lref\BilalES{
A.~Bilal and S.~Metzger,
``Anomaly cancellation in M-theory: A critical review,''
Nucl.\ Phys.\ B {\bf 675}, 416 (2003)
[arXiv:hep-th/0307152].
}

\lref\mooretalk{An online version is available at http://online.kitp.ucsb.edu/online/mp03/moore1. } 

 \lref\dsfone{D.~S.~Freed, ``Dirac charge quantization and
generalized differential cohomology,'' Differ. Geom. VII (2000) 129--194
[arXiv:hep-th/0011220].
} 

 \lref\dsftwo{D.~S.~Freed, ``$K$-theory in quantum field theory,'' Current
Developments in Mathematics 2001, International Press, Somerville, MA,
41-87
[arXiv:math-ph/0206031].
}

\lref\EvslinKN{
J.~Evslin,
``From E(8) to F via T,''
arXiv:hep-th/0311235.
}

\lref\MathaiMU{
V.~Mathai and H.~Sati,
``Some Relations between Twisted K-theory and E8 Gauge Theory,''
arXiv:hep-th/0312033.
}

\lref\dupont{J. Dupont, R. Hain\ and\ S. Zucker, Regulators and characteristic classes
    of flat bundles, in {\it The arithmetic and geometry
    of algebraic cycles (Banff, AB, 1998)}, 47--92, Amer. Math. Soc.,
    Providence, RI, 2000, alg-geom/9202023}
\lref\freedcsii{D. S. Freed, Classical Chern-Simons Theory II, Houston J. Math. {\bf 28}
    (2002), no.~2, 293--310}
 
\lref\chernsimons{  S. S. Chern and J. Simons, ``  Characteristic forms and geometric
    invariants,''  Ann. Math.  {\bf 99 }(  1974)  pp.  48--69}




\Title{\vbox{\baselineskip12pt \hbox{hep-th/0312069}
\hbox{RUNHETC-2003-34}
}}%
{\vbox{\centerline{The $M$-theory $3$-form  and $E_8$ gauge theory } }}

\smallskip
\centerline{Emanuel Diaconescu \&  Gregory Moore}
\medskip

\centerline{\it Department of Physics, Rutgers University}
\centerline{\it Piscataway, New Jersey, 08855-0849}

\bigskip

\centerline{and}

\bigskip

\centerline{Daniel S. Freed}
\medskip
\centerline{\it Department of Mathematics  }
\centerline{\it University of Texas at Austin}

\smallskip
\bigskip 
\noindent
We give a precise formulation of the $M$-theory $3$-form potential $C$ in a
fashion applicable to topologically nontrivial situations.  In our model the
$3$-form is related to the Chern-Simons form of an $E_8$ gauge field.  This
leads to a precise version of the Chern-Simons interaction of 11-dimensional
supergravity on manifolds with and without boundary.  As an application of
the formalism we give a formula for the electric $C$-field charge, as an
integral cohomology class, induced by self-interactions of the $3$-form and
by gravity. As further applications, we identify the $M$ theory Chern-Simons
term as a cubic refinement of a trilinear form, we clarify the physical
nature of Witten's global anomaly for 5-brane partition functions, we clarify
the relation of $M$-theory flux quantization to $K$-theoretic quantization of
RR charge, and we indicate how the formalism can be applied to heterotic
$M$-theory.

\bigskip
\noindent 

\Date{Nov. 30, 2003; Revised March 22, 2004}


\def\adp{ { {\rm ad} P} } 

\newsec{ Introduction }

This  paper summarizes a talk given at the conference on 
Elliptic Cohomology at the Isaac Newton Institute, in December, 2002 
\mooretalk. 

In this paper we will discuss the relation of $M$-theory  
to $E_8$ gauge theory in  10,11, and 12 dimensions. 
Our basic philosophy is that 
formulating $M$-theory in a mathematically precise way, 
in the   presence of nontrivial topology, 
challenges our understanding of the fundamental 
formulation of the theory, and therefore might lead to 
a deeper understanding of how one should express 
the unified theory of which $11$-dimensional supergravity 
and the five $10$-dimensional string theories are 
distinct limits.  
To be more specific, let us formulate   three motivating problems
for the formalism we will develop.

The first problem concerns $11$-dimensional supergravity. 
We will be considering physics on an $11$-dimensional, 
oriented, spin  manifold, $Y$. 
When it has a boundary we will denote $\p Y = X$. 
The basic fields of $11$-dimensional supergravity are a
 metric $g $, (Lorentzian or Euclidean) 
 a ``$C$-field,''  and a gravitino $\psi \in \Gamma(S\otimes T^*Y) $
where $S$ is the spin bundle on $Y$. 
Our main concern in this paper is with the mathematical nature 
of the $C$-field. In the standard formulation of supergravity 
the $C$-field is regarded as   a  3-form gauge potential 
\eqn\cefl{
C \in \Omega^3(Y). 
}
This generalizes the 
Maxwell potential $A\in \Omega^1(M) $ of electromagnetism on a 
manifold $M$.  The fieldstrength is defined to be
\eqn\fldstr{
G = dC \in \Omega^4(Y).
}
%
Whereas in {\it classical\/} electromagnetism the gauge potential $A$ is
a global 1-form, in the {\it quantum\/} theory Dirac's law of charge
quantization changes the geometric nature of~$A$: it is now a connection on a
$U(1)$ line bundle over $M$.  (See \dsftwo\ (sec. 3) and \dsfone\ (sec.2) for
expository 
accounts of how charge quantization leads to a $U(1)$~connection and its
generalizations in differential cohomology theory.) 
One of our goals is to give a similar geometric description 
of the $C$-field. 

In standard supergravity the path integral measure is a canonical formal
measure on the space of fields, defined by the metric $g$ and the 
Faddeev-Popov procedure, weighted by an   action. 
Schematically, the exponentiated action is given by 
%
\eqn\sugraact{
\exp\Biggl[ -2\pi \int_Y {1\over  \l^9} \vol(g) \CR(g)+ {1\over 2\l^3} G\wedge *G  + \bar \psi \Dsl \psi \Biggr] 
\Phi(C)
}
plus $4$-fermion terms, 
where $\l$ is the $11$-dimensional Planck length and, roughly speaking, 
\eqn\navphi{ \Phi(C) \sim   
\exp\biggl(2\pi i \int_Y {1\over 6} C G^2 - C I_8(g) \biggr) 
}
Here $I_8(g)$ is a quartic polynomial in the curvature tensor. 
We will give precise normalizations for the $C$-field and $I_8$ in sections
3.2 and 4.1 below. 
 
Now let us suppose we wish to formulate the action in the presence of 
nontrivial topology. This means, among other things, that we wish to allow 
the fieldstrength $G$ in \fldstr\ to define a nontrivial class in the DeRham 
cohomology of $Y$. Evidentally, we cannot use \navphi. One might be tempted to 
introduce a $12$-manifold bounding $Y$ and use Stokes formula. This procedure 
works (after accounting for several subtleties) when $Y$ is closed but fails 
when $\p Y \not=\emptyset$. Thus our  first problem is: 
{\it Find a mathematically precise definition of $\Phi(C)$ when $G$ is 
cohomologically nontrivial, and $\p Y = X$ is nonempty. } 
We will give a complete answer to this question in section 5 below.

Having formulated the measure we can next turn to applications. 
 When $\p Y = X$ is nonempty, the path integral for  the $C$-field on a manifold 
with boundary $Y$ defines a wavefunction of the boundary 
values $C_X$ of $C$. We may denote this wavefunction as 
$\Psi(C_X)$. Now, there is a group of gauge transformations $\CG$ of the 
$C$-field and the wavefunction must be suitably 
gauge invariant.  Our second problem is {\it If $\Psi(C_X)$ is a nonvanishing 
gauge invariant wavefunction, what conditions are imposed on the 
values of $C_X$? } Put more simply: {\it What is the Gauss law 
for the $C$-field? } We will find nontrivial conditions on $C_X$ 
in section 7.1 and will interpret them as the condition that 
the induced electric $C$-field charge associated to a $C$-field 
configuration and gravity must vanish. 

 An analogy might be useful at this point. 
Because of the Chern-Simons phase $\Phi(C)$, the path integral is a 
generalization of the familiar  3-dimensional massive  gauge theory: 
\eqn\csgt{
\Psi(A_X) = \int [d A_Y] \exp\biggl[-\int {1\over 2 e^2} {\Tr} F\wedge *F + 
 i {k\over 4\pi} \int_Y {\Tr} \biggl(AdA + {2\over 3} A^3\biggr)\biggr] 
}
where $A_Y$ is a connection on a bundle $P_Y$ over the three-manifold $Y$
and $A_X$ is the fixed boundary value. 
The ``Gauss law''  is the statement that this ``function'' is suitably 
gauge invariant. 
 The Gauss law implies, among other things, that one can only define 
a nonvanishing 
gauge invariant wavefunction when $c_1(P_X)=0$, 
where $P_X$ is the restriction of $P_Y$.  In the 
case $e^2 = \infty  $ we have pure Chern-Simons theory.  
The Gauss law then implies   $F(A_X)=0$ and   leads directly to the 
mathematical interpretation of $\Psi$ as a section of a complex line bundle 
over the moduli space of flat connections on a Riemann surface.  

The analogy to Chern-Simons gauge theory cannot be pushed too far. 
In \csgt\ the integer parameter $k$ appears. In \navphi\ the analogous 
parameter is, roughly speaking, $k=1/6$. It is precisely this fractional 
value which makes the proper definition of $\Phi(C)$ subtle. 

Finally, and most importantly, 
 let us turn to the third motivating problem, namely an understanding of 
the   emergence of $E_8$ {\it gauge} symmetry in $M$-theory. 
There are several hints that there is a fundamental role for the group $E_8$ in 
$M$-theory. First, there are the famous $U$-duality global symmetries that
arise when $M$-theory is compactified on tori. These symmetries involve 
exceptional groups.  Next the duality relating 
$M$ theory on $K3$ to heterotic string theory on $T^3$ implies that there are 
enhanced  gauge symmetries when the $K3$-surface develops $A-D-E$ singularities.  
These gauge symmetries certainly include $E_8 \times E_8$. 
Next, a  construction of Horava-Witten shows that 
a quotient of $M$-theory on $X \times S^1$ by an 
orientation reversing isometry of $S^1$ leads to 
the $E_8 \times E_8 $ heterotic string, with $E_8$ gauge fields propagating 
on the boundary \HoravaMA\HoravaQA.
Furthermore, in \WittenMD\  Witten gave a definition of $\Phi(C)$ that used 
$E_8$ gauge theory in 12 dimensions. 
This definition was then used in \DiaconescuWY\   to establish a 
connection to the $K$-theoretic classification of 
RR fluxes in the limit that $M$ theory reduces to 
type II string theory. 
All this suggests a hidden $E_8$ structure in $M$ theory 
which might point the way to a useful reformulation of the 
theory.  Thus we have our third problem: 
{\it What is the precise relation of the $C$-field of 
11-dimensional supergravity to an $E_8$ gauge field? } 
We will propose an answer to this question in section 3. 
The remainder of the paper endeavors to demonstrate that 
this answer can be useful.

Finally, we would like to close this introduction with a general remark. 
In the physics literature gauge fields are traditionally modeled as
belonging to a space with a group action, or simply as an element of the
quotient space.  For example, a 1-form gauge field is typically taken to be a
connection on a fixed principal bundle, and the group of gauge
transformations acts on the space of all such connections.  It is technically
sounder to use a different model.  Namely, we consider instead the {\it
groupoid\/} of all connections on all principal bundles; morphisms are maps
of principal bundles which preserve the connections.  There is still a
quotient space of equivalence classes of connections, and this is the space
over which one writes the functional integral.  This model is local, whereas
fixing a particular principal bundle is not.  Also, we can replace the
groupoid by an equivalent groupoid without changing the physics.  This allows
us the convenience of using 
 different models adapted to different purposes.   
In section 3 we introduce a few different models for the
$C$~field. One (described in secs. 3.1-3.3) is closer to the physics tradition, 
while the other two (described in secs. 3.4-3.5) involves a  groupoid model.  
In this paper we mostly rely on the first model, which 
emphasizes the connection to~$E_8$.  There is an alternative approach to some
of the issues in this paper \freedhopkins\ which uses yet other models,
following the ideas developed in \HopkinsRD.  That point of view leans more
heavily on the (differential) algebraic topology to construct the cubic form
which appears in the M-theory action.  In particular, the model for the
$C$-field does not involve $E_8$ gauge fields.

Two very recent papers on subjects closely related to the present work 
are \EvslinKN\MathaiMU.

\newsec{The gauge equivalence class of a  $C$-field } 

Our first task is to give a precise answer to the question: 
``What is a $C$-field?'' 
In this section we will give a partial answer, by describing 
the gauge equivalence class (gec) of a $C$-field. The answer will be 
that  the  gec of a $C$-field is a 
(shifted) differential character.

To motivate this description let us consider the 
description of the gauge equivalence class of a 
 $U(1)$ gauge field 
on a manifold $M$. The key to answering this problem 
turns out to be to consider  the holonomy around $1$-cycles $\gamma$. 
This is certainly gauge-invariant information, and it turns out 
to be all the gauge invariant information. More precisely, 
given a connection $A$ on a line bundle over $M$ we 
may regard the holonomy as a map on closed 1-cycles 
$\c\chi_A: Z_1(M) \to U(1)$ given by 
\eqn\holon{
\c \chi_A(\gamma) = \exp\biggl(i \oint_{\gamma} A \biggr)
} 
It is then natural to ask how   $\c \chi_A$ differs from an arbitrary such map. 
The answer is that there exists a closed 2-form, the {\it fieldstrength of $\c \chi$},
 such that if $\gamma = \p B_2$ is a boundary then 
\eqn\stks{
\c \chi_A(\gamma) = \exp\biggl(i \int_{B_2} F \biggr)
}
Note that it follows that $F$ has   $2\pi \IZ$-periods, denoted
$F\in \Omega^2_{2\pi \IZ}(M)$, and that $F$ is closed. 
Note also that it follows from \stks\ that $\c \chi$ is a homomorphism of abelian 
groups. It is smooth in an appropriate sense. 
Such maps $\c \chi$ such that a fieldstrength exist  are in 1-1 correspondence 
with the gec  of $U(1)$ bundles with connection. 

In supergravity theories one often encounters $p$-form gauge potentials. One 
natural mathematically precise formulation of such gauge potentials is in 
terms of {\it differential characters}, which generalize the above description 
of $U(1)$ gauge connections. By definition a   {\it Cheeger-Simons character},
 or {\it differential character} 
$\c \chi\in \c H^{p+1}(M)$ 
is a   homomorphism 
\eqn\dficar{
\c \chi: Z_p(M) \to U(1)
}
where $Z_p(M)$ is the group of $p$-cycles on $M$, 
such that there is a {\it fieldstrength} $\omega(\c \chi) \in \Omega^{p+1}_{\IZ}(M)$. 
That is, there exists a closed differential $(p+1)$-form $\omega(\c \chi) $ 
with $\IZ$-periods such that 
\eqn\fldsk{
\Sigma_p = \p B_{p+1}\quad \Rightarrow \quad 
\c \chi(\Sigma_p ) = \exp\biggl(2\pi i \int_{B_{p+1}} \omega(\c \chi)  \biggr).
}

We will identify the space of gec's  of a $p$-form gaugefield with the space of 
differential characters $\c \chi \in \c H^{p+1}(M)$. In order to establish 
some notation let us recall the basic facts about differential characters. 
(See \HopkinsRD\ and references therein for further details about differential 
characters and cohomology). 

The gauge invariant information in a Cheeger-Simons character can 
be expressed in two distinct ways, each of which is summarized by 
an exact sequence. The first sequence is related to the space of 
{\it flat characters,} that is, characters with   $\omega(\c \chi)=0$: 
\eqn\extsqi{ 
0 \rightarrow H^p(M,U(1)) \to \check{H}^{p+1}(M) \to \Omega^{p+1}_{  \IZ}(M) \rightarrow 0 
}
This is clear because a flat character defines a homomorphism of 
abelian groups $H_p(M,\IZ) \to U(1)$, which, by Poincar\'e duality 
is $H^p(M,U(1))$. 

In order to define the second sequence we begin with the 
{\it topologically trivial characters}. If  
 $A\in \Omega^p(M)$ is globally defined, then we may define a differential character:  
\eqn\toptrivc{
\c \chi_{A}(\Sigma_p) := \exp[2\pi i \int_{\Sigma_p} A]. 
}
Note that, first of all,  $\c \chi_A$ only depends on $A$ modulo $\Omega^p_{\IZ}$, 
and secondly, the fieldstrength  $\omega(\c \chi_A) = dA$ 
is trivial in cohomology. Thus, the cohomology class of the fieldstrength 
is an obstruction to writing a character as a trivial character \toptrivc. 
In fact we have the second sequence: 
\eqn\extsqii{
0 \rightarrow \Omega^p/\Omega^{p}_{\IZ} \to \check{H}^{p+1}(M) \to H^{p+1}(M,\IZ) \rightarrow 0 
}
The  projection map in this sequence defines the {\it characteristic class} of $\c \chi$
which we will denote as $a(\c \chi) \in H^{p+1}(M,\IZ).$ Note that this is a class in 
integral cohomology, not DeRham cohomology. The compatibility relation 
between the two sequences  states that 
\eqn\compat{
a(\c \chi)_{\IR} = [\omega(\c \chi)]_{DR}
}
where $a(\c \chi)_{\IR}$ denotes the image of $a(\c \chi)$ in DeRham cohomology.

It turns out that the gec of a $C$-field is not quite a differential character, but 
is rather a ``shifted'' differential character \WittenMD. The reason for this is best explained in 
terms of the coupling to the ``membrane.'' In the formulation of $M$-theory - as it is 
presently understood - one posits the existence of fundamental ``electrically charged'' 
membranes with 3-dimensional worldvolumes. 
  In Maxwell theory, a charged particle, of charge $e$,  moving 
along a worldline   $\gamma$, couples to the background gauge potential $A$ 
via the holonomy: 
\eqn\maxcoupl{\exp[ i \int_{\gamma} e A ] .
}
In $M$ theory the membrane couples  to the $C$-field in an 
analogous way. The standard coupling of supergravity fields to 
the membrane wrapping a 3-cycle $\Sigma$ is usually written: 
\eqn\memhol{
\sim 
\exp[ 2\pi i \int_{\Sigma} C ] .
}
More accurately, because of the worldvolume fermions on the membrane, the topologically 
interesting part of the membrane amplitude is 
\eqn\membcoup{
\sqrt{\Det \Dsl_{S(\CN)} } \exp\biggl(2\pi i \int_{\Sigma} C \biggr)
} 
where $\CN$ is the rank 8 normal bundle of the embedding $\iota: \Sigma \hookrightarrow Y$ 
and $S(\CN)$ is an associated chiral spinor bundle. The squareroot has an ambiguous 
sign when one considers \membcoup\ on the space of all 3-cycles $Z_3(Y)$ and metrics, 
and the holonomy \memhol\ has a compensating sign such that the product is well-defined 
\WittenMD. The net effect is that it is the gec of the 
difference $ [C_1 - C_2]$ of $C$-fields which is an honest differential character in $ \check H^4(Y)$. 
We can be slightly more precise about the nature of the shift.  
As shown in \WittenMD\ the    requirement that 
\membcoup\ is well-defined implies quantization condition 
on  the   fieldstrength:
\eqn\geflq{
[G]_{DR} = a_{\IR} - \half \lambda_{\IR}  
}
  where 
$a\in H^4(Y,\IZ)$ is an integral class and $\lambda $  is the canonical integral 
class of  the spin bundle of  $Y$. 
%
%
\foot{Put differently, only twice the characteristic class of the twisted differential character 
is well-defined, and it is constrained to be equal to $w_4$ modulo two. } 
 
In conclusion the we have a partial answer to the question ``What is a $C$-field?''  
The gauge equivalence class of a $C$-field is a shifted differential character. 
We have not yet defined precisely what is meant by  ``shifted.'' 
This will be explained in the section 3.4 (equation 3.25).  The space of shifted characters, with 
shift $\half \lambda$ will be denoted $\c H^4_{\half \lambda}(Y)$.

\newsec{ Models for the $C$-field }

In the previous section 
we have explained what the  gauge equivalence class of a $C$-field 
is, but have {\it not}  answered the question: `` What is a $C$-field?''
An analogous situation would be to have in hand a formulation of nonabelian gauge theory  
in terms of gauge invariant quantities without having introduced connections on 
principal bundles. In physical theories the requirement of locality, that we 
be able to formulate  the 
theory in terms of local fields,   forces us to introduce redundant variables,  
such as gauge potentials. Similarly, in the standard formulation of supergravity 
one takes the $C$ field to be an ordinary 3-form  $C \in \Omega^3(Y)$ subject to a
gauge invariance $C \to C + \omega$ where $\omega$ is a closed 3-form.  
When  $\omega$ is 
exact such gauge transformations
are referred to as {\it small gauge transformations.}  When $\omega$ is closed but 
not exact the gauge transformation is a {\it large gauge transformation}. 
In $M$-theory, quantization 
of membrane charge requires that $\omega$ have integral periods. 
In this view, the gauge equivalence class of a $C$-field is 
\eqn\naves{ [C]\in \Omega^3(Y)/\Omega^3_{\IZ}(Y).
}
According to \extsqii\ such 
fields define topologically trivial differential characters. However,  
 many  interesting nontrivial phenomena in string/$M$-theory involve 
topologically nontrivial characters and hence we must modify the geometric 
description of the $C$-field. In this paper we will focus on  
the ``$E_8$ model for the $C$-field,'' which seems well-suited to 
describing the $M$-theory action. As we will discuss below, there are 
other models of the $C$-field which can be considered to be equivalent
to the $E_8$ model.

\subsec{The  $E_8$ model for the $C$-field }

The $E_8$ model is motivated  
by Witten's definition \WittenMD\ of the $M$-theory action  
as an integral in 12-dimensions. (We will review Witten's definition 
in section 4.1 below.) This model is based on the topological fact that 
there is   a homotopy equivalence 
\eqn\homtequiv{
\eqalign{
E_8 & \sim K(\IZ,3) \cr}
}
up to the 14-skeleton. Equivalently, the homotopy groups $\pi_i(E_8)$ of $E_8$ vanish 
for $4\leq i \leq 14$. It follows from 
\homtequiv\ that $BE_8   \sim K(\IZ,4)$ and therefore,   for  
 $\dim M\leq 15$, there is a one-one 
correspondence between integral classes 
\eqn\intgrlcs{
a\in H^4(M,\IZ)
}
and isomorphism classes of principal $E_8$ bundles over $M$. 
\foot{A more elementary way to understand this is to use the 
obstruction theory arguments of \WittenBT.} 
For each $a$ we pick a specific bundle $P(a) \to M$. 
\foot{This is somewhat unnatural and will lead to problems 
with gluing of manifolds and hence with locality. 
 One can easily modify the 
definition below to include triples $(P,A,c)$ where $P$ is 
an $E_8$ bundle in the isomorphism class determined by $a$. 
The discussion that follows is not changed in any essential 
way, except that one must account for bundle isomorphisms 
when discussing equivalences of $(P,A,c)$ with $(P',A',c')$. 
We have suppressed this refinement here in the interest of 
simplicity and brevity.} 
We now come to a central definition;  we will 
say that a   {\it ``$C$-field  on $Y$  with characteristic class $a$ }''  is an element of 
\eqn\eespace{
\E_P(Y) := \CA(P(a)) \times \Omega^3(Y)
}
where $\CA(P)$ is the space of smooth connections on the 
principal bundle $P$. 
Thus,   our   ``gauge potentials,'' or ``$C$-fields,'' 
will be pairs $ (A,c)\in \E_P(Y)$.  
We will often denote $C$-fields  by $\c C = (A,c)$, 
and we will   call $c$ the ``little $c$-field.''

\subsec{
The gauge equivalence class of $\c C$ }

Given the ``gauge potential''  $\c C = (A,c)\in \E_P(Y)$  we will now describe its 
gauge equivalence class. The principle we use is that the holonomy, or  coupling of the $C$-field to 
an elementary membrane contains all the gauge invariant information in $C$.  
The holonomy of  $\c C = (A,c) \in \E_P(Y) $ around $\Sigma$ is defined to be 
\eqn\ceehol{
\c \chi_{A,c}(\Sigma) := 
\exp\Biggl[ 2\pi i \biggl( \int_{\Sigma} CS(A) -{1\over 2} CS(g)+ c \biggr) \Biggr] 
}
Here $CS(A)$ and $CS(g)$ are Chern-Simons invariants associated to the gauge field $A$
and the metric $g$,  normalized by 
\foot{In general $CS(A)$ is not well defined as a differential form, and only 
$\exp[2\pi i \int_{\Sigma} CS(A)]$ is well-defined. See  \chernsimons. For a recent account 
see \freedcsii, sec.1.2. 
Below we will have 
occasion to use the relative Chern-Simons invariant $CS(A,A') :=\int_{[0,1]} \tr F^2 $ 
defined by integrating along a straightline between the two connections. 
This is well-defined as a differential form.} 
%
%
\eqn\norml{ 
d CS(A) = {\tr } F^2 :=  {1\over 60} {\Tr}_{\bf 248}\bigl({F^2\over 8 \pi^2}\bigr)  
}
\eqn\normlai{ 
d CS(g) = {\tr} R^2:= -{1\over 16 \pi^2} {\Tr}_{\bf 11} R^2
}
Note that 
\eqn\normliii{ 
[{\tr} F^2]_{DR} = a_{\IR} \qquad [{\tr} R^2]_{DR} = \half (p_1(TY))_{\IR} 
}
It follows immediately that the fieldstrength of the character is 
\eqn\fldstrn{ 
\omega(\c \chi_{A,c})= 
G = {\tr } F^2 - {1\over 2} {\tr} R^2 + dc 
}

Note that the   normalizations are chosen
 such that     $\exp[2\pi i \int_{\Sigma} CS(A)]$   is well-defined. 
 However,     $\exp[ \pi i \int_{\Sigma} CS(g)]$  has a sign ambiguity when regarded as a function 
on $Z_3(Y) \times Met(Y)$. This sign ambiguity is cancelled by the sign ambiguity of 
the worldvolume fermion determinant: 
\eqn\dfsn{
\sqrt{\Det(\Dsl_{S(\CN)}) } 
}
It is the metric dependence required to cancel the sign ambiguity of \dfsn\ which 
leads to $\c \chi_{A,c}$ being a shifted character.

 Previous authors have proposed that 
there should be an identification of the
 $C$ field with an $E_8$ Chern-Simons form.  See, for examples, \DiaconescuWZ,
 and \morrison.  However, a formula such as $C = CS(A) $ is
 unsatisfactory for many reasons.  A similar idea based on $OSp(1\vert 32)$
 was proposed in \horava.  Related ideas, in which the $3$-form gauge
 potential of 11-dimensional supergravity should be considered as a composite
 field, have been explored in \MeloschWM.  The model we have just described,
( which was inspired by the $\IR/\IZ$ index theory of Lott \lott, and was
 first announced in \MooreNN)   fits into this circle of ideas, but we emphasize that the
 $E_8$ connection plays a purely auxiliary role, at least on manifolds 
without (spatial) boundaries.   This becomes more clear when
 one considers alternative formulations which do not involve $E_8$ at all.

%
%

\subsec{The $C$-field   gauge group } 

Now that we have defined ``gauge potentials''  
 we seek a gauge group $\CG$ so that 
\eqn\ggi{
\E_P(Y)/\CG =  \c H^4_{\half \lambda}(Y).
}
The   fiber of the group orbit should be defined by 
\eqn\fibr{ 
(A,c) \sim (A',c') \qquad \Leftrightarrow \qquad \c \chi_{A,c} = \c \chi_{A',c'} 
}
This condition is easily solved: $(A,c) \sim (A',c')$ iff there exists 
$\alpha \in \Omega^1(\adp) $ and $\omega\in \Omega^3_{\IZ}(Y)$  such that : 
\eqn\coni{
A' = A+ \alpha \qquad \alpha \in \Omega^1(\adp) 
}
\eqn\cnii{
 c' = c- CS(A,A+\alpha) + \omega \qquad \omega\in \Omega^3_{\IZ}(Y)
}
To prove this, note that equality of fieldstrengths implies 
$c'-c = CS(A',A) + \omega$  for some closed $3$-form $\omega$. 
Then equality of the holonomies shows that $\omega$ must have 
integral periods.

From \fibr\coni\cnii\  we might conclude that the ``$C$-field gauge group'' is: 
\eqn\smlgg{
\CG ~{\buildrel ? \over = } ~ \Omega^1(\adp) \sdtimes \Omega^3_{\IZ}(Y) .
}
The right hand side of \smlgg\  is indeed a group, with group law
\eqn\grpsil{
(\alpha_1, \omega_1) (\alpha_2, \omega_2) 
= \left(\alpha_1 + \alpha_2, \omega_1 + \omega_2 + d({\tr}\alpha_2\wedge \alpha_1)\right)
}
and does satisfy \ggi. Nevertheless, it is {\it not} precisely the gauge group we need.  
This can be understood in two ways, one physical and one mathematical.

Let us first explain the physical point of view.  
In electromagnetism, if $\p Y = X$, the worldline of a 
charge particle can end on a point $P\in X$. The coupling 
of the charged particle to the background gauge potential
\eqn\coupl{ \exp[i \int_{\gamma}e A ] 
}
is not gauge invariant. Now note that gauge transformations in Maxwell theory
can be thought of as elements 
$\c \chi \in \c H^1(Y)$, since $\c H^1(Y)$ is just the group of $U(1)$-valued 
functions on $Y$. In this view, the gauge transformation by $\c \chi$ 
of the  ``open wilson line'' 
\coupl\ is: 
\eqn\oepns{
 \exp[i \int_{\gamma}e A ] \to \c \chi(P) 
\exp[i \int_{\gamma}e A ] .
}

  In $M$ theory, the worldvolume of a membrane $\Sigma$ can 
end on a 2-cycle $\sigma$. By analogy with electromagnetism we should define 
 $C$-field gauge transformations to act on such ``open membrane Wilson lines''    as: 

\eqn\openmdm{
\exp[2\pi i \int_{\Sigma} C] \to \c \chi(\sigma) \exp[2\pi i \int_{\Sigma} C] 
}
where $\exp[2\pi i \int_{\Sigma} C] $ is short for \ceehol, and  
 $\c \chi(\sigma)$    is a 
$U(1)$-valued function on $Z_2(X)$. Moreover, the existence of a fieldstrength for 
$C$ shows that there must exist a fieldstrength for $\c \chi$. Indeed, 
suppose  $\sigma = \p \Sigma$, and $C \to C+ x$ with $x\in \Omega^3_{\IZ}(Y)$. 
Then, if $\Sigma \subset X$,  
consistency with \openmdm\ demands $\c\chi(\sigma) = \exp[2\pi i \int_{\Sigma} \omega] $.
But this is the defining property of $\c \chi\in \c H^3(X)$ !
Since we could choose any 10-dimensional subspace $X$ in $Y$ in this discussion, 
we conclude that   the gauge transformation parameter  should be regarded as  $\c \chi\in \c H^3(Y)$.
In this way we conclude that the proper definition of the gauge group should be 
\eqn\propdf{
 \CG := \Omega^1(\adp ) \sdtimes \c H^3(Y)
}
Recall that 
\eqn\cndtsn{ 
0 \to H^2(Y,U(1)) \to \c H^3(Y) \to \Omega^3_{\IZ}(Y) \to 0 
}
and thus we have a nontrivial extension of the naive gauge group \smlgg. 

The gauge group \propdf\ acts on $\E_P(Y)$ via  
\eqn\ggfrpact{
(\alpha, \c \chi)\cdot (A,c) = \left(A + \alpha, c- CS(A, A+ \alpha) + \omega(\c \chi)\right)
}
The group law is 
\eqn\newgplw{
(\alpha_1, \c \chi_1) (\alpha_2, \c \chi_2) = 
(\alpha_1 + \alpha_2, \c \chi_1 + \c \chi_2 + \c \chi_{b} ) 
}
where $\c \chi_b$ is a topologically trivial character with 
$b = {\Tr} (\alpha_2 \alpha_1)$.  

It is convenient to introduce some terminology. We will refer to 
gauge transformations in the connected component of the 
identity in \propdf\ as {\it small gauge transformations}. 
Otherwise, the gauge transformation is referred to as a {\it large 
gauge transformation.} We will refer to gauge transformations in 
$H^2(Y,U(1))$ as {\it micro gauge transformations} since they leave 
$(A,c)$ unchanged. 
\foot{Note that some micro gauge transformations can also be large
gauge transformations! The characteristic class $a(\c \chi)$ for such 
gauge transformations will be torsion.} 
As we shall see, they nevertheless have a crucial 
physical effect. 
 
This formalism makes clear the physical role of the $E_8$ gauge field. It is 
a kind of topological field theory since we can shift $A$   
to {\it any} other connection $A' \in \CA(P(a))$, and hence $A$ is only constrained 
by topology. 
\foot{In Donaldson-Witten theory there is also a symmetry under arbitrary 
shifts of the gauge potential, 
  $\delta A = \psi$. In that case $\psi$ is nilpotent. }

\subsec{$C$-fields and categories }

Let us now turn to a more  mathematical justification of the definition
\propdf. 
Quite generally, bosonic fields with internal symmetry should be viewed as
objects of a {\it groupoid\/}---a category with all arrows invertible---and
equivalent groupoids provide alternative models for the same physical
object.  In this section we demonstrate an equivalence of our model with
another possible model for the $C$-field. 

To begin, since we have a $\CG$-action on a space $\E_P(Y)$ we can form an 
associated groupoid. We   regard this groupoid as a category. 
The objects of this category are elements of $ \E_P(Y)$. 
The morphisms $Mor(\c C_1, \c C_2)$ are the group elements taking 
$\c C_1$ to $\c C_2$. We denote this category by 
$\E_P(Y)//\CG$.  Note from \ggfrpact\ that the 
 objects $\c C$ in the category have automorphism group given by 
the flat characters of degree three: $H^2(Y,U(1))$. 

Now, in  \HopkinsRD\ M. Hopkins and I. Singer have formulated a 
theory of   ``differential cochains'' and ``differential cocycles'' 
which refines the theory of differential cohomology. 
Differential cocycles  may be regarded as one definition of 
what is meant by  gauge potentials  
 for abelian $p$-form gauge fields.\foot{%
For some abelian $p$-form gauge fields, such as the Ramond-Ramond fields in
Type~II superstring theory, the quantization law is not in terms of {\it
ordinary\/} cohomology so gauge fields are defined in terms of ``differential
functions'' and ``generalized differential cohomology theories.''}
In the  framework of \HopkinsRD\  a ``shifted 
differential character'' is the equivalence class of
a differential cocycle which trivializes a specific 
differential 5-cocycle related to $W_5(Y)$.  
The Hopkins-Singer  theory can therefore be applied to the 
$C$-field of $M$-theory.

To be more specific,  
\foot{This paragraph assumes some familiarity with \HopkinsRD.} 
the cohomology class $w_4(Y)\in H^4(Y,\IZ_2)$ defines a 
differential cohomology class,  $\c w_4$, because 
we can include $H^4(Y,\IZ_2) \hookrightarrow H^4(Y, \IR/\IZ)
\hookrightarrow \c H^4(Y)$. 
The characteristic class of the flat character $\c w_4 $ is 
the integral class $W_5(Y)\in H^5(Y,\IZ)$, given by 
the Bockstein homomorphism applied to $w_4(Y)$. This class is interpreted 
as the background magnetic charge induced by the topology 
of $Y$, and this class must vanish in order to be able to 
formulate any (electric) $C$-field at all.  
On a spin manifold, $W_5(Y)=0$, since the   class 
$\lambda$ is an integral lift of $w_4(Y)$. 
When $W_5(Y)=0$  we 
may refine the differential cohomology class $\c w_4$ to a 
differential cocycle by defining 
\eqn\canfive{ 
\c W_5 = (0 , \half \lambda(g) , 0) \in \c Z^5(Y) \subset
C^5(Y, \Z) \times C^4(Y, \R) \times \Omega^5(Y)
}
where $\lambda(g) = -{1\over 16 \pi^2}{\Tr}R^2$ is functorially attached
to the metric $g$. 
We then define a $C$-field to be a differential cochain 
$\c C = (\bar a,h, \omega) \in C^4(Y,\IZ) \times C^3(Y,\IR) \times \Omega^4(Y)$ 
trivializing $\c W_5$:
\eqn\tirvals{
\delta \c C = \c W_5. 
}
%
Written out explicitly this means that
\eqn\triplesc{
\eqalign{
\delta \bar a= 0 , \cr
\delta h = \omega- \bar a_{\R} + \half \lambda(g), \cr
d\omega =0. \cr}
}
%
We refer to these as shifted differential cocycles, and denote the 
space of such cochains by   $\c Z^4_{\half \lambda(g)} $. This is a 
principal homogeneous space for $\c Z^4(Y)$ and hence   may be 
considered as the set of objects in   a category.

Now, finally, we may explain  the mathematical motivation 
for the choice of  gauge group \propdf. It is only 
with this choice that  we have the crucial theorem 

\bgn
\bgn 
{\bf Theorem. } There is an equivalence of  the  categories $\E_P(Y)//\CG$
and $\c Z^4_{\half \lambda(g)} $.

\bigskip
{\it Proof}: 
To prove this  it suffices to establish the existence of a fully faithful 
functor $F: \E_P(Y)//\CG  \to \c Z^4_{\half \lambda(g)}$ such that any 
object in $\c Z^4_{\half \lambda(g)}$ is isomorphic to $F(\c C)$ for 
some $\c C$. Since both categories are groupoids, the morphism 
spaces are principal homogeneous spaces for the automorphism group. 
Since the automorphism group is independent of object and category, 
namely $H^2(Y,U(1))$, and since set of   isomorphism classes 
of objects is the same, namely $\c H^4_{\half \lambda}(Y)$ it simply 
suffices to establish the existence of a functor.

Begin by choosing a Hopkins-Singer
 $4$-cocycle $(\bar c_0, h_0, \omega_0)$ on the classifying space 
$BE_8$ where $[\bar c_0]$ is a generator of $H^4(BE_8;\IZ)$, 
and $\omega_0$ is determined from the universal connection 
$A_{univ} $ on $EE_8$ by $\omega_0 = \tr F(A_{univ})^2$. 
Now, there exists a map $\gamma: P \to EE_8$ which classifies the
  connection: $\gamma^*(A_{univ}) = A$.
\foot{For a nice proof, see \dupont, section 2.  One needn't rely on this,
  however.  Instead, we introduce an equivalent category which includes a
  classifying map; see \freedcsii, section 3.1.}
%
Let $\bar \gamma:Y \to BE_8$ be 
the induced classifying map on the base space Then we define our functor 
on objects by:  
$F(A,c) = (\bar c_{HS}, h_{HS}, \omega_{HS}) $ where 
\eqn\functordf{
\eqalign{
\bar c_{HS} & = \bar \gamma^*(c_0)   \cr
h_{HS} & = \bar \gamma^*( h_0) +c   \cr
\omega_{HS} & = \bar \gamma^*(\omega_0)+ dc = {\tr} F^2(A) + dc -\half \lambda(g) \cr}
}
and one checks that $(\bar c_{HS}, h_{HS},\omega_{HS} )$ is a 
shifted differential cocycle. 
$\spadesuit$

In particular, and crucially, the 
automorphism group of a $C$-field (analogous to the constant gauge 
transformations in nonabelian gauge theory) is the same in both 
categories, namely $H^2(Y,U(1))$.

Finally, given the $E_8$ model for the $C$-field a natural 
question one may ask  is how $E_8$ gauge transformations, 
that is, bundle automorphisms of $P$, are related to   
the $C$-field gauge transformations. The answer involves a
construction which will prove useful later. Suppose 
\eqn\bdlaut{
(A,c) \to (A^g, c) \qquad g\in Aut(P)
}
is an $E_8$ gauge transformation. 
Every transformation of this type is equivalent to a $C$-field gauge 
transformation $(\alpha,\c \chi)$ acting on $(A,c)$. We first find $\alpha$, trivially
it is 
\eqn\bdlauti{
\alpha = A^g - A := g^{-1} D_A g .
}
It follows that $\omega(\c \chi) = CS(A, A+\alpha) = CS(A,A^g)$. 
A natural way to construct a character with this fieldstrength
 is to use $g$ to construct the 
twisted bundle over $X \times S^1$
\eqn\twisbdl{
P_g : = (P \times [0,1])/(p,0) \sim (pg,1)
}
and take the vertical connection $dt \p_t + d_X + \CA $
where 
\eqn\curlya{
\CA = (1-t) A + t A^g
}
We can then set $\c \chi= \c \chi_{(g,A)}$ where 
\eqn\chiga{ 
\c \chi_{(g,A)}(\sigma) := \exp\Biggl[ 2\pi i \int_{\sigma \times S^1} CS(\CA) \Biggr] 
}
It is straightforward to check that
\eqn\fieldsch{
\omega(\c \chi_{(g,A)})= CS(A,A^g) = -{1\over 3} {\tr} (g^{-1} D_A g)^3 + db(g,A)
}
where $b(g,A)$ is globally well-defined. 
It is rather interesting to note that $Aut(P)$ is {\it not} a subgroup of $\CG$, 
since $(\alpha, \c \chi)$ depends on $A$. 
Of course, since it is a group acting on $\E_P(X)$ it does define a 
{\it sub-groupoid. } 
We will come back to this point in section 12 below.

%
%
%

\subsec{A third model for the $C$-field}


There is a different approach to differential cohomology theory
based on differential function spaces \HopkinsRD.\ This motivates a 
different model for the 
$C$-field which will be described in some detail in the following. 
Of particular importance is the filtration on differential function spaces.
While this approach will be less familiar to physicists, because  
gauge transformations are replaced by morphisms in a category,  
the approach has an appealing flexibility.

As motivation for the construction, let us begin 
with a simpler example, namely the group of differential 
$2$-characters ${\check H}^2(Y)$, which is isomorphic to the group of
equivalence classes of line bundles with connection on $Y$. 

One can construct a cocycle category for 
this group as follows. We take the objects to be pairs $(L,A)$ 
consisting of a line bundle with connection on $Y$. The space of 
morphisms between two objects $(L,A)$, $(L',A')$ consists of 
equivalence classes of pairs $(\CL,\CA)$ on $Y\times \Delta^1$
so that\foot{%
More precisely, a morphism is a pair~$(\CL,\CA)$ together with isomorphisms $(3.32)$
all up to the equivalence relation below.  Also, note that the equivalence
relation on morphisms and the composition of morphisms can be defined by
working on~$Y\times \Delta ^2$, where $\Delta ^2$~is the standard 2-simplex.}
\eqn\morphA{ 
(\CL,\CA)|_{Y\times \{0\}} = (L,A)\qquad  
(\CL,\CA)|_{Y\times \{1\}} = (L',A').}
Here $\Delta^1$ is the standard 1-simplex. 
Two pairs $(\CL_0,\CA_0)$, $(\CL_1,\CA_1)$ satisfying the same boundary 
conditions \morphA\ are said to be equivalent if they are homotopy equivalent 
relative to the boundary conditions. This means that there exists a 
pair  
$({\bf L}, {\bf A})$ on $Y\times \Delta^1 \times I$ which restricts 
to $(\CL_i,\CA_i)$ on $Y\times \Delta^1 \times \{i\}$, for $i=0,1$. 
Moreover, $({\bf L}, {\bf A})$ should also restrict to $(L,A)$, $(L',A')$ 
on the two boundary components $Y\times (\partial \Delta^1)\times I$. 
Composition of morphisms is defined by concatenation of paths. 

\def\ra{\rightarrow}

One can check that the construction sketched above yields a groupoid, but it 
is easy to see that this is not a cocycle category for ${\check H}^2(Y)$.
Two objects $(L,A)$, $(L',A')$ are equivalent if they are connected by a morphism 
$(\CL,\CA)$. Then the parallel transport associated to the connection 
$\CA$ determines an isomorphism $\phi:L\ra L'$. For a general connection $\CA$, 
$\phi$ is not compatible with the connections $A, A'$. Therefore the equivalence classes 
of objects are in one to one correspondence to line bundles on $Y$ up to isomorphism, 
which is not the answer we want. 

In order to obtain ${\check H}^2(Y)$ in this manner we have to refine our groupoid structure 
so that the resulting equivalence relations are compatible with connections. This is a special case 
of a general construction described in detail in \HopkinsRD.\ The main idea is to impose an extra 
condition on morphisms by keeping only pairs $(\CL,\CA)$ so that 
\eqn\morphB{ 
F^{1,1}(\CA)=0}
where $F^{1,1}(\CA)$ denotes the component of the curvature with one leg along $\Delta^1$. 
The effect of this condition is that the parallel transport along $\Delta^1$ defined by $\CA$ 
becomes horizontal. Then the isomorphism $\phi:L\ra L'$ preserves connections $\phi^*A'=A$, and 
we obtain the desired set of equivalence classes. Note that for consistency we have to impose 
a similar filtering condition on the pairs $({\bf L}, {\bf A})$ introduced below \morphA.\ 
Namely, we require all components with legs along $I\times \Delta^1$ to vanish 
\eqn\morphC{ 
F^{1,1}({\bf A}) = F^{0,2}({\bf A})=0.} 
Note that this construction is valid even if $Y$ is a manifold with boundary since we did 
not have to invoke integration by parts at any stage. The same will be true for the category of 
$C$-fields constructed below. (In physical applications one might wish to impose 
boundary conditions.) 

In the following we will produce a cocycle  category $\CE$ 
for $C$-fields on a manifold $Y$ proceeding in a similar way. 
The objects are triples $(P,A,c)$ as in section 3.1. Hence $P$ is a principal $E_8$ bundle on $Y$, 
$A$ is a connection on $P$ and $c\in \Omega^3(Y)$. The ``curvature'' of a
triple is the closed  4-form $G$ defined by
\eqn\fldstr{G = {\tr } F^2 - {1\over 2} {\tr} R^2 + dc}
The space of morphisms between two objects 
$(P,A,c)$ and $(P',A',c')$ consists of equivalence classes of triples
$(\CP,\CA,\gamma)$ 
on $Y\times \Delta^1$ so that 
\eqn\cfieldA{ 
(\CP, \CA, \gamma)|_{Y\times \{0\}}=(P,A,c) \qquad 
(\CP, \CA, \gamma)|_{Y\times \{1\}}=(P',A',c').}
Denoting the curvature of $(\CP,\CA, \gamma)$ by $\CG$, we  
 impose a filtering condition  $\CG^{3,1}=0$,
where $\CG^{(4-k,k)}$, $k=0,1$ denotes the component of $\CG$ 
of degree $k$ along $\Delta^1$. Since $d\CG=0$, this implies that $\CG$ is pulled back 
from $Y$. (This means that $G$ is ``gauge invariant,'' as it should be.) 
 Two triples $(\CP_0,\CA_0,\gamma_0)$, $(\CP_1,\CA_1,\gamma_1)$ satisfying the 
same boundary conditions \cfieldA\ are said to be equivalent if they are homotopy 
equivalent relative to the boundary. This means that there exists a triple 
$({\bf P}, {\bf A}, {\bf c})$ on $Y\times \Delta^1\times I$ with obvious restriction 
properties, as explained below \morphA.\ For consistency we have to impose a 
filtering condition ${\bf G}^{3,1}={\bf G}^{2,2}=0$ analogous to \morphC.\ 
One can check that this construction defines a groupoid $\CE$.

We claim this 
is a correct cocycle category for the $C$-field. In the remaining part of this 
section we will verify this claim by showing that 

$i)$ the equivalence classes of objects 
are in one to one correspondence with shifted differential characters, and 

$ii)$ the automorphism group of any object $(P,A,c)$ is isomorphic to
$H^2(Y,\IR/\IZ)$.  

Let us start with $(i)$.  
Two objects $(P,A,c)$ and $(P',A,',c')$ are equivalent if they are connected by a morphism. 
Suppose $(\CP,\CA,\gamma)$ represents such a morphism. Since $\CG$ is pulled back from $Y$,
using the boundary conditions \cfieldA\ we find that $\CG=G=G'$. Moreover, the parallel transport 
associated to the connection $\CA$ determines an isomorphism $\phi':P\ra P'$. Therefore 
$P,P'$ have the same characteristic class $a\in H^4(Y,\IZ)$. We conclude that we have a 
well defined map 
\eqn\cfieldB{ 
{\widetilde \CE} \ra A^4_{\half\lambda}(Y)}
where ${\widetilde \CE}$ denotes the set of equivalence classes of objects of
$\CE$ and
$A^4_{{1\over 2}\lambda}(Y)$ denotes the set of pairs 
$(a,G)\in H^4(Y,\IZ) \times \Omega_{\IZ+{\lambda\over 2}}^4(Y)$ subject to 
the compatibility condition $a_\IR-{\lambda_\IR\over 2} = [G]_{DR}$. 
Note that 
$A^4_{{1\over 2}\lambda}(Y)$ is a principal homogeneous space over the 
group $A^4(Y)$ of pairs $(a,G)$ subject 
to the unshifted compatibility condition $a_\IR=[G]_{DR}$. 
This map is clearly surjective. This is an encouraging sign since the group of shifted 
differential characters ${\check H}_{{1\over 2}\lambda}$ is expected to surject onto this space. 
In order to finish the identification, we should show that the fiber of this map is isomorphic 
to the torus $H^3(Y,\IR)/H^3(Y,\IZ)$. 

The fiber over a point $(a,G)\in A^4_{\half\lambda}(Y)$ consists of isomorphism classes 
of triples $(P,A,c)$ with fixed $(a,G)$. Up to isomorphism, a triple satisfying this condition
can always be taken of the form $(P_0,A_0,c)$ for some fixed $(P_0,A_0)$ with $a(P_0)=a$. 
Therefore it suffices to determine the set of isomorphism classes of triples of 
the form $(P_0,A_0,c)$ with fixed $(a,G)$. Since only the three-form $c$ is allowed to vary, 
it is straightforward to check that these triples are parameterized by the space of 
closed three-forms $z\in \Omega^3_{cl}(Y)$. Two triples $(P_0,A_0,c), (P_0,A_0,c+z)$ are isomorphic 
if they are connected by a morphism $(\CP,\CA,\gamma)$ in the groupoid.  
Now we make use of the filtering condition $\CG^{3,1}=0$, which yields 
\eqn\cfieldC{ 
d_t\gamma^{(3,0)} +d_Y\gamma^{(2,1)} + ({\rm tr} F(\CA)^2)^{(3,1)}=0.}
Integrating this relation along $\Delta^1$, and using the boundary conditions \cfieldA,\ 
we find 
\eqn\cfieldD{ 
z = -d_Y \beta -\eta\qquad \beta = \int_{ \Delta^1} \gamma\qquad 
\eta = \int_{ \Delta^1} ({\rm tr} F(\CA)^2).}
At this point note that the bundle with connection $(\CP,\CA)$ on $Y\times \Delta^1$ 
satisfies identical boundary conditions along the two boundary components $Y\times \{0\}$ 
and $Y\times \{1\}$. By a standard gluing argument, we can construct a 
bundle with connection $({\widetilde \CP}, 
{\widetilde \CA})$ over $Y\times S^1$. Then we have 
\eqn\cfieldE{ 
\eta=\int_{ S^1} ({\rm tr} F({\widetilde \CA})^2),}
which is a closed form with integral periods. 
In conclusion, the three-forms $c$ and $c+z$ parameterizing isomorphic triples $(P_0,A_0,c)$, 
$(P_0,A_0,c+z)$ with fixed 
$(a,G)$ differ by closed forms with integral periods. However $\Omega^3_{cl}(Y)/\Omega^3_{\IZ}(Y)
\simeq H^3(Y,\IR)/H^3(Y,\IZ)$, hence we obtain the expected result. 
This shows that the gauge equivalence classes of $C$-fields in this model are indeed shifted 
differential characters. 

To conclude this section, let us compute the automorphism group of an arbitrary object $(P,A,c)$ 
of $\CE$. Using the boundary conditions \cfieldA\ and a standard gluing argument, we can easily show 
that $\hbox{Aut}(P,A,c)$ consists of equivalence classes of triples $(\CP,\CA,\gamma)$ on $Y\times S^1$ 
subject to the condition $\CG^{3,1}=0$. Moreover, $S^1$ is equipped with a base point $*\in S^1$, and 
$(\CP,\CA,\gamma)$ restricts to  $(P,A,c)$ on $Y\times \{*\}$. As noted below \cfieldA,\ since $\CG$ 
is closed, it follows that it is a pull back from $Y$, $\CG = \pi^* G$, where $\pi:Y\times S^1\ra Y$ 
is the canonical projection, and $G$ is the curvature of $(P,A,c)$.

An important observation is that triples $(\CP,\CA,\gamma)$ on $Y\times S^1$
 satisfying $\CG^{3,1}=0$ form themselves 
a groupoid $\CM$. 
The morphism space between two such triples $(\CP,\CA,\gamma)$ and $(\CP',\CA',\gamma')$ consists of triples 
$({\bf P}, {\bf A}, {\bf c})$ on $Y\times S^1\times \Delta^1$ which restrict to 
$(\CP,\CA,\gamma)$ and respectively $(\CP',\CA',\gamma')$ on the two components of the boundary. 
Two triples $({\bf P}_0, {\bf A}_0, {\bf c}_0)$ and $({\bf P}_1, {\bf A}_1, {\bf c}_1)$
are said to be equivalent if they are homotopy equivalent relative to boundary 
conditions. Furthermore, we have to impose a filtering condition of the form ${\bf G}^{3,1}=
{\bf G}^{2,2}=0$, where ${\bf G}^{k,4-k}$ denotes the component of ${\bf G}$ of degree $k$ along $Y$. 
Note that $\CM$ is in fact a cocycle category for $C$-fields on $Y\times S^1$ subject to a 
filtering condition on the curvature. 

The automorphisms of $(P,A,c)$ are classified by equivalence classes of objects of $\CM$ 
so that $(\CP,\CA,\gamma)|_{Y\times \{*\}} = (P,A,c)$. Proceeding by analogy with $\CE$, the 
equivalence classes of objects of $\CM$ are in one to one correspondence to differential 
characters ${\check \chi}\in {\check H}^4(Y\times S^1)$ so that $\omega({\check \chi})^{3,1}=0$. 
In particular, this implies that $\omega({\check \chi})$ is pulled back from $Y$. 
Adding the extra condition  $(\CP,\CA,\gamma)|_{Y\times \{*\}} = (P,A,c)$ fixes 
${\check \chi}|_{Y\times \{*\}} ={\check \rho}$, where $\c \rho\in {\check H}^4(Y)$ is 
the character determined by $(P,A,c)$. 
Using the exact sequence 
\eqn\cfieldG{ 
0\ra H^3(Y\times S^1, \IR/\IZ) \ra {\check H}^4(Y\times S^1) \ra \Omega^4_{\IZ}(Y\times S^1) \ra 0}
and the K\"unneth formula $H^3(Y\times S^1,\IR/\IZ)\simeq H^3(Y,\IR/\IZ) \oplus H^2(Y,\IR/\IZ)$, 
we see that $\c \rho$ fixes the component in the first summand, but not the second. 
It follows that the  characters $(\CP,\CA,\gamma)$ are parametrized by $H^2(Y,\IR/\IZ)$. 
In conclusion, $\hbox{Aut}(P,A,c)\simeq H^2(Y,\IR/\IZ)$ for any object $(P,A,c)$. This is 
in agreement with our previous discussion in section 3.3. 

We have so far given two different constructions of cocycle groupoids for $C$-fields which 
have identical equivalence classes of objects and automorphism groups. Therefore the two groupoids 
must be equivalent.

\newsec{ The definition of the $C$-field measure for $Y$ without boundary }

\subsec{Witten's definition}

In \WittenMD\ Witten gave a definition of the phase factor $\Phi(C)$ 
in \sugraact\   using $E_8$ gauge  theory in $12$-dimensions.  We 
will review his definition, and then show how to recast it in 
intrinsically 11-dimensional terms. The 11-dimensional formulation 
will then be in a form suitable to generalization to the case when 
$Y$ has a boundary.

Suppose $P(a) \to Y$ admits an extension 
\eqn\twfl{
 P_Z(a_Z ) \to Z}
where   $Z $ is a bounding spin manifold (and hence 
$\p Y = \emptyset$). The existence of such an 
extension follows from Stong's theorem \stong.
Suppose, moreover, that  $\c C_Z = (A_Z, c_Z)$ 
extends $\c C_Y= (A_Y, c_Y) \in \E_P(Y)$ to $\E_{P_Z(a_Z)}(Z)$ 
(such extensions always exist if $P_Z(a_Z)$ exists).  Then 
Witten's definition is 
\eqn\wittdef{ 
\Phi_{\rm W}(\c C_Y; Y) = \exp\Biggl\{  2\pi i \int_Z \bigl[{1\over 6} G_Z^3 - G_Z I_8(g_Z)\bigr] 
\Biggr\} 
}
where 
\eqn\geefieldex{ 
G_Z =   {\tr} F_Z^2 - {1\over 2} {\tr} R_Z^2 + dc_Z 
}
and $I_8(g)$ is defined on a manifold $M$ by 
\eqn\iate{
I_8(g) =  {1\over (4\pi)^4} \biggl(\bigl({\Tr} R^2\bigr)^2 - 4{\Tr} R^4\biggr).
}
$I_8(g)$ 
is a closed form, depending on a metric, which represents 
\eqn\iateii{
I_8 := {1\over 48 } (p_2(TM) - \lambda(TM)^2) 
}
in $H^8(M, \IR)$, where we recall that $\lambda = p_1/2$. 

One needs to check that \wittdef\ is independent of the 
extensions $a_Z, P_Z, \c C_Z$. This is almost true 
thanks to some remarkable identities involving $E_8$ 
index theory. To check independence of extensions it suffices 
to check that the right hand side
 of \wittdef\ is equal to 1 on a closed 
12-manifold $Z$. 
\foot{It is quite crucial that we extend the entire triple $(A_Y, c_Y, P_Y)$, 
and not just the fieldstrength $G_Z$. Otherwise $\Phi$ can vary 
continuously with a choice of extension.   }

Consider the Dirac operator $\Dsl_{V(a_Z)}$ coupled to 
the bundle $V(a_Z)$ associated to $P_Z(a_Z)$ via the 
${\bf 248}$ of $E_8$, and endowed with connection 
$A_Z$. The index density is formed from 
\eqn\indxdsn{
i(\Dsl_{V(a_Z)}):={\Tr}_{\bf 248}\exp[ F_Z/2\pi] \hat A(g_Z)
}
where $F_Z$ is the fieldstrength of $A_Z$, 
while $\hat A(g_Z)$ is the usual Dirac index density formed from the 
curvature 2-form $R$, computed using $g_Z$. 
 Using properties 
of $E_8$ we find that 
\eqn\denti{
{\Tr}_{\bf 248}\exp\left( F_Z/2\pi\right) = 248 + 60 \alpha_Z + 6\alpha_Z^2 + {1\over 3}\alpha_Z^3
}
where $\alpha_Z= {1\over 60} {\Tr}_{\bf 248} F^2/8\pi^2$ represents 
$(a_Z)_{\IR}$ in DeRham cohomology. 
Now, let $i(\Dsl_{RS})$ denote the 12-dimensional gravitino
index density
\foot{This is not the gravitino density of a $12$-dimensional theory, but rather that 
appropriate to an $11$-dimensional theory.}  $\hat A {\Tr}\bigl( e^{R/2\pi} -1 \bigr) + 8 \hat A$. 
Then, extracting the 12-form part, we have the crucial identity
\eqn\indexdens{
\eqalign{
& 
\biggl[
\half i(\Dsl_A) + {1\over 4} i(\Dsl_{RS}) 
\biggr]^{(12)} = {1\over 6} G_Z^3 - G_Z I_8(g_Z)\cr
& - d\Biggl[ c_Z (\half G_Z^2 - I_8(g_Z))-\half c_Z d c_Z G_Z +{1\over 6} c_Z (d c_Z)^2
\Biggr]  \cr}
}
and thus, on a closed $12$-manifold $Z$ the RHS of \wittdef\ is 
\eqn\wittrhs{
\exp\left\{ 2\pi i \left(\half I(\Dsl_{V(a_Z)}) + {1\over 4} I(\Dsl_{RS})\right) \right\}
}
where $I$ is the index. In 12-dimensions the index is always even, 
and hence \wittrhs\ simplifies to a sign $(-1)^{I(\Dsl_{RS})/2}$. 
Thus, \wittdef\ is independent of extension, up to the sign $(-1)^{I(\Dsl_{RS})/2}$.

\subsec{Intrinsically 11-dimensional definition} 

We would now like to recast the definition \wittdef\ into intrinsically 11-dimensional 
terms. This is readily done using the APS index theorem: 
\eqn\apsicn{
I(\Dsl_E) = \int_Z i(\Dsl_E) - \xi(\Dsl_E)
}
for the Dirac operator coupled to a bundle with connection $E$. 
Here  
\eqn\etfan{ 
\xi(\Dsl_E) := \half( \eta(\Dsl_E) + h(\Dsl_E) ) 
}
This is the standard invariant involving the $\eta$ invariant and the 
number of zeromodes $h(\Dsl_E)$. 

Motivated by \wittdef, \indexdens, and \apsicn\ we {\it define}
\eqn\precisecs{
\Phi(\c C;Y) :=  \exp
\biggl[ i \pi \xi(\Dsl_{V(a)}) + {i \pi \over 2} \xi(\Dsl_{RS}) + 2\pi i I_{\rm local} \biggr] 
}
The total derivative in \indexdens\ leads to the elementary local factor 
\eqn\locfac{ 
I_{\rm local} =  
 \int_Y\left\{ c \bigl( {1\over 2}G^2 -I_8(g)\bigr) -
{1\over 2}cdc  G +{1\over 6}c(dc)^2\right\}
}
The expression $\xi(\Dsl_{RS})$   is short for 
\eqn\rsfactr{
 \xi(\Dsl_{T^*Y } ) - 3 \xi(\Dsl_{ } )   
}
%

 As a function of 
the $C$-field, \precisecs\  is gauge invariant, that is, 
$\Phi(A,c; Y) = \Phi(A',c';Y) $
for $(A',c') = (\alpha,\c\chi) \cdot (A,c)$. Thus it is a $U(1)$-valued 
function on $\c H^4_{\half \lambda}(Y)$.

Let us now comment briefly on  $\Phi$ as a function of the 
metric $g$.  Note that we have 
dropped a sign-factor  in passing from 
\wittdef\ to \precisecs: 
\eqn\diffr{
\Phi_{\rm W} = (-1)^{I(\Dsl_{RS})/2} \Phi
}
Let $\CD$ be the locus of metrics $g$ such that $\Dsl_{T^*Y}$ and $\Dsl$ have 
zero modes, and suppose $g\in Met(Y)-\CD$.
 On $Met(Y)-\CD$ ,   $\Phi$ is well-defined as a number.
 However, it does not 
extend continuously to the full space of metrics $Met(Y)$. 
This difficulty was essentially solved in \WittenMD. 
One should consider $\Phi$ as a continuous section 
of a complex line bundle associated to a principal $\IZ_2$ 
bundle $T$.  Then the   gravitino path integral, which we will 
denote schematically as ${\rm Pfaff}(\Dsl_{RS})$ is a section of 
a real line bundle associated to a canonically isomorphic 
principal $\IZ_2$ bundle, 
and will have (in general) a first order zero along the locus 
of zeromodes. 
\foot{The definition of this principal $\IZ_2$ bundle and the canonical 
isomorphism will be explained in detail in \toappear.}
 The product 
${\rm Pfaff}(\Dsl_{RS})\cdot  \Phi$ becomes a well-defined smooth
diffeomorphism-invariant  function on 
the gauge equivalence classes of metrics and $C$-fields. 
   
There is a standard formal measure $\mu(\c C)$ for $3$-form 
gauge potentials defined by the the metric $g$ on $Y$ together with 
the  Faddeev-Popov procedure applied 
to small $C$-field gauge transformations. Taking this into account, 
the well-defined measure for the $C$-field path integral, after integrating out 
the gravitino, is 
\eqn\ceemeasure{
\mu(\c C) e^{-{1\over 2\l^3} \int_Y G\wedge *G} {\rm Pfaff}(\Dsl_{RS}) \cdot 
\Phi(\c C;Y) .
}

\newsec{The  $C$-field measure when $Y$ has a boundary }

Now we are ready to reap some of the benefits of the $E_8$ 
model for the $C$-field. Using this formalism we will 
completely solve problem 1 of the introduction.

When we try to formulate the $C$-field 
measure in the   case where $\p Y = X $ is nonempty {\it the same formula}
\precisecs\ holds. In particular we continue to take 
\eqn\phibdry{ 
\Phi(\c C;Y) =  \exp
\biggl[ i \pi \xi(\Dsl_A) + {i \pi \over 2} \xi(\Dsl_{RS}) + 2\pi i I_{\rm local} \biggr] 
}
However,   there is now  a conceptually important distinction: 
the factor $\exp[i \pi \xi(\Dsl_A)]$ is a section of a $U(1)$ 
bundle with connection over the space of $C$-fields on $X$.
 As in the case when $Y$ has no 
boundary, the  factor $\exp[\half i \pi \xi(\Dsl_{RS} )]$
is more subtle and must be combined with the gravitino determinant. 
We hope to discuss this   elsewhere \toappear. 
In this paper we consider a fixed metric $g \in Met(Y)-\CD$.


It is well-known in  Chern-Simons theory that in the presence 
of a boundary the exponentiated invariant $\exp[i \pi \xi(\Dsl_E)]$ 
should be viewed as a section of a line bundle. 
In our case  $\Phi$ is   valued in 
a principal $U(1)$ bundle 
\eqn\quebudl{
\CQ \to \E_P(X).
}
This bundle can be defined by the following property: 
Each extension $\c C_Y $ of $\c C_X\in \E_P(X)$, defined for 
some $Y$ with $\p Y = X$,  produces an 
element:
\eqn\bdlele{
\Phi(\c C_Y;Y) \in \CQ_{\c C_X }
}
and two such extensions satisfy the ``gluing law'' 
\eqn\gluelaw{
{\Phi(\c C_{Y}; Y) \over \Phi(\c C_{Y'} ; Y') } = 
\Phi(\c C_Y - \c C_{Y'} ; Y  \bar Y') 
}
where $\bar Y$ denotes orientation reversal. 
Alternatively,  
we may define $\CQ_{\c C_X}$ more directly as follows. 
Suppose $P\to X$ is a fixed $E_8$ bundle of 
characteristic class $a_X$. For any   extension 
$P_Y$ of $P$   define 
$\E_{P_Y}(\c C_X)$ to be the set of all extensions $\c C_Y $ of $\c C_X$. 
\foot{Since $\Omega^{\rm spin}_{10}(pt) = \IZ_2 \oplus \IZ_2\oplus \IZ_2$ 
and $\tilde \Omega^{\rm spin}_{10}(K(\IZ,4)) = \IZ_2 \oplus \IZ_2$ in general 
we cannot extend $P\to X$. However, in the physical situations of 
interest there will always be such an extension since $X$ is the boundary 
of the theory defined on the spacetime $Y$.} 
We may then define the fiber $\CQ_{\c C_X}$ to be the set of all $U(1)$-valued 
functions $F$ on $\E_{P_Y}(\c C_X)$, for some $Y$, 
 such that the ratio $F(\c C_Y)/F(\c C_{Y'})$ 
is the right hand side of \gluelaw. This space of functions is a principal 
homogenous space for $U(1)$ and varies smoothly with $\c C_X$.   
Thus the fibers define  a smooth $U(1)$ bundle $\CQ \to \E_P(X)$.


 The circle bundle $\CQ$ we have just defined carries a canonical 
connection. Recall that a connection is simply a law for 
  lifting of paths in the base space to the total space satisfying 
the natural composition laws. 
A path $p(t) = (A_X(t),c_X(t))$, $0 \leq t \leq T$ in $\E_P(X)$ 
defines a $C$-field on the cylinder $  \c C_p \in \E_P(X \times [0,T])$. 
So 
\eqn\canconn{
\Phi(\c C_p; X \times [0,T]) \in \CQ_{(A,c)(T)} \otimes \CQ_{(A,c)(0)}^* 
= {\rm Hom}(\CQ_{(A,c)(0)},\CQ_{(A,c)(T)})
}
defines the appropriate parallel transport.

Given the explicit formula \phibdry\ we can give an explicit formula for 
the connection. Choose an extension $(Y, P_Y)$ of $(X,P)$. 
Since every cotangent vector $(\delta A, \delta c) $ to $\E_P(X)$ has an extension to 
a cotangent vector of $\E_{P_Y}$ 
it suffices to evaluate the covariant derivative on a section provided by 
an infinitesimal family of  extensions $\c C_Y$. The main variational formula is: 
\eqn\variationalp{
{\nabla \Phi(\c C_Y;Y) \over \Phi(\c C_Y;Y)} = 
2\pi i \int_Y \bigl( \half G^2 - I_8\bigr) \bigl( 2 {\tr}(\delta A F) + \delta c\bigr) 
}
where 
$2 {\tr}(\delta A F) + \delta c $ is a form of type $(1,3)$ on $E_P(Y) \times Y$. 
 
It is now a simple matter to compute the curvature.
Identifying
 the tangent space $T \E_P(X) = \Omega^1(ad P) \oplus \Omega^3(X)$ 
evaluation on tangent vectors $(\alpha_i, \chi_i) $ gives the curvature 
$2$-form $\Omega$ of $\CQ$: 
\eqn\curvaturei{
\Omega((\alpha_1, \xi_1);(\alpha_2, \xi_2)) = 2 \pi i \int_X G \wedge \bigl( 2 {\tr}(\alpha_1 F ) + \xi_1 \bigr)
\wedge  \bigl( 2 {\tr}(\alpha_2 F ) + \xi_2 \bigr)
}

   The theory   of determinant line bundles shows that the 
associated line bundle $\CL$ to $\CQ$ may be regarded as a 
determinant line bundle. More precisely,  $\CL = {\rm PFAFF}(\Dsl_{\adp})$, 
the Pfaffian line bundle for adjoint $E_8$ fermions on the boundary $X$.  
But, we emphasize, we are not introducing physical $E_8$ fermions on the 
boundary. 
 
The line bundle we have defined can also be defined by considering 
families of   Dirac operators with boundary conditions along the lines 
of  \DaiKQ. This definition fits in more naturally with the inclusion of the 
Pfaffian of the gravitino operator and will be discussed in \toappear. 
It is also possible to formulate the line, together with its connection, in the 
categorical approach of sections 3.4 and 3.5.

\newsec{ The action of the gauge group on the physical wavefunction and the Gauss law } 
 
In this section we will study the action of the $C$-field gauge group
on the wavefunctions defined by the $C$-field path integral. 
Fix an extension $P_Y$ of $P$ to $Y$. For $\c C_X \in \E_P(X)$ recall that 
$\E_{P_Y}(\c C_X)$ is  the set of all extensions $\c C_Y $ of $\c C_X$. 
Then we set, formally, 
\eqn\frmlwve{ 
\Psi(\c C_X) = \int_{\E_{P_Y}( \c C_X)/\CG(\c C_X) } 
\mu(\c C_Y) ~ \exp[-{1\over 2\l^3} \int G \wedge * G ] ~ \Phi(\c C_Y; Y) 
}
Here $\CG(\c C_X)$ is the group of $C$-field gauge transformations on 
$Y$ which fixes $\c C_X$. 
That is, we integrate over all isomorphism classes of extensions $\c C_Y$  of $\c C_X$.  

As we have seen, $\Phi(\c C_Y;Y)$ is valued in $\CQ_{\c C_X}$ and 
hence $\Psi(\c C_X)$ is valued in the    line associated to $\CQ_{\c C_X}$.
We will denote this line $\CL_{\c C_X}$.

\subsec{The Gauss Law }

Physically meaningful wavefunctions must be gauge invariant. 
In our case, the wavefunction is a section of a line bundle 
$\CL \to \E_P(X)$. In order to formulate gauge invariance we 
will need to define a lift of the group action: 
\eqn\lifted{
\matrix{
\CL& \qquad & {\buildrel \CG\over \rightarrow} & \qquad \CL \cr
    &        &                                  &     \cr
\downarrow &  &  &  \downarrow \cr
     &  &  & \cr
 \E_P(X) & \qquad & {\buildrel \CG\over \rightarrow} & \qquad  \E_P(X) \cr}
}
Then the condition of gauge invariance is simply
\eqn\gausslaw{
g \cdot \Psi(\c C_X) = \Psi(g\cdot \c C_X)
}
for all $g\in \CG, \c C_X\in \E_P(X)$. 
This condition is known as the {\it Gauss law}. 

Formally, any wavefunction defined by a path integral with 
gauge invariant measure, such as \frmlwve\ automatically 
defines a gauge invariant wavefunction. This is essentially 
because the path integral integrates over gauge copies of 
fields and thus projects onto gauge invariant quantities. 
 Of course, this projection might vanish, so the Gauss law 
should be considered, in part,  as a condition on $\c C_X$ such that 
nonvanishing gauge invariant wavefunctions can be supported on 
$\c C_X$. 

We  will now give a precise 
definition of the group lift \lifted. We will then use that 
to derive an important consequence of the Gauss law.

\subsec{ Definition of the $\CG$ action on $\CL$}

The main result of this section is the construction of 
 a well-defined lift \lifted. 
The full construction is rather lengthy. We confine ourselves to
sketching  the 
most important part, namely defining the action of gauge transformations of the 
form $(0,\c \chi)$ on $\CL$.  
We will construct the group lift using the natural connection 
on $\CL$ defined above. To do this, we introduce some 
standard   paths in $\E_P(X)$. For any $\xi \in \Omega^3(X)$ 
define   the linear path 
\eqn\linpth{
p_{\c C, \xi}(t) = \bigl(A, c+ t \xi \bigr) 
}
connecting $\c C$ to $\c C + \xi$. (It is   convenient 
to write $\c C + \xi := (A, c+ \xi)$ for a $3$-form $\xi$. )
 
Now, if $\Psi \in \CL_{\c C}$ and $\c \chi\in \c H^3(X)$ then we will define 
%
\eqn\defgplf{
(0,\c \chi)\cdot \Psi := \varphi(\c C, \c \chi)^* U(p_{\c C, \omega(\c \chi)}) \cdot \Psi
}
where $\varphi$ is a phase. 
The reason we must introduce the phase factor $\varphi$ in \defgplf\ is 
that the connection on $\CL$ has curvature. 
Indeed, an elementary computation shows that we have: 
\eqn\curvcomp{
U(p_{\c C,\omega( \c \chi_1+ \c \chi_2)}) = U(p_{\c \chi_1 \c C, \omega(\c \chi_2)})
 U(p_{\c C, \omega(\c \chi_1)}) 
e^{i \pi \int_X G \wedge \omega(\c \chi_1) \wedge \omega(\c \chi_2) } 
}
where $\c\chi_1\c C$ denotes the gauge transform of $\c C$ by $\c \chi_1$. 

 It follows that to define a group theory lift we will need to find functions 
$\varphi(\c C, \c \chi)$   such that: 
\eqn\cocyrel{
\varphi(\c C, \c \chi_1+ \c \chi_2) = \varphi(\c \chi_1 \c C, \c \chi_2) \varphi(\c C, \c \chi_1) 
e^{+i \pi \int_X G \wedge \omega(\c \chi_1) \wedge \omega(\c \chi_2) } .
}

We now construct a cocycle satisfying the   
relation \cocyrel.
%
%
( In the following section we will provide some physical motivation for this 
choice of $\varphi$. )
We use the data $\c C_X$ and $\c \chi$ to construct the    
character on the {\it  closed}  11-fold $Y= X \times S^1$
defined by 
\eqn\twischar{
\pi_1^*[\c C_X] + \pi_2^*(\c t )\cdot \pi_1^*(\c \chi)
}
Here $\pi_i$ are projections (henceforth dropped) and   $\c t\in \c H^1(S^1)$ is 
the canonical character associated with $S^1 \cong U(1)$.
 This character \twischar\  has fieldstrength $G + dt \wedge \omega(\c \chi)$, 
characteristic class $a + [dt]   a(\c \chi)$ and holonomy 
\eqn\holoi{
\c\chi_{(A,c)}(\Sigma_3) e^{2\pi i t \int_{\Sigma_3}\omega(\c \chi) }
}
on cycles of type $\Sigma_3 \times \{ t \}$ and holonomy 
$\c \chi(\sigma_2) $ on cycles of type $\sigma_2 \times S^1$. 
We will refer to the character \twischar\ as the twist of 
$[\c C_X]$ by $\c \chi$. 
Choosing the Neveu-Schwarz, or bounding spin structure, $S^1_-$, we define 
the correction factor in terms of the   the $M$-theory 
phase \precisecs: 
\eqn\mphasecf{
\varphi(\c C_X, \c \chi):= \Phi( [\c C_X] + \c t \cdot \c \chi; X \times S^1_-) 
}
Note that since  $\Phi$ is defined as a function  on $\c H^4_{\half \lambda}(Y)$ 
for closed spin 11-manifolds $Y$ equation \mphasecf\ makes sense. Also note 
that, since we are working with a fixed metric a definite choice of the sign 
of $\exp[i \pi \xi(\Dsl_{RS})/2]$ must be made here.

The idea behind \mphasecf\   is essentially 
stated in \WittenHC,  footnote 19. 
An easy argument shows that \mphasecf\ indeed satisfies the requisite cocycle law. 
The cocycle \mphasecf\  can be expressed in terms of $\eta$ invariants of Dirac operators 
by using the construction of equations \bdlaut\ - \fieldsch\ above. 
A similar construction allows one to define the group lift for the entire 
group $\CG$. 

We need to check that  \mphasecf\ really satisfies \cocyrel. 
The proof   is a 
standard cobordism argument. We consider a differential 
character on  the spin 11-manifold 
$(X\times S^1_-) \cup (X\times S^1_-) \cup (X\times S^1_-) $
restricting to $[\check C ]+  \check t\cdot\check \chi_1, 
[\check C ]+  \check t\cdot\check \chi_2,
[\check C ]-\check t\cdot( \check \chi_1+\check \chi_2)
$
on the three components. We then choose the extending 
spin 12 manifold to be $Z=  X\times \Delta $ where 
$\Delta$ is a pair of pants bounding the three 
circles with spin structure restricting 
to $S^1_-$ on the three components.  To be explicit we can choose 
  $\Delta $ to be  the simplex 
$\{ (t_1, t_2): 0\leq t_1\leq t_2 \leq 1 \}$
with identifications $t_i \sim t_i + 1$. Now 
we extend 
the differential character as 
\eqn\extndchr{
[\c C] + \c t_1 \cdot \c \chi_1 + \c t_2 \cdot \c \chi_2 
}
which clearly restricts to the required character on 
the boundary. The field strength is  
$ G_Z = G_X +  dt_1\wedge \omega_1  +  dt_2\wedge \omega_2$. 
For such extensions, $\Phi_{\rm Witten} = \Phi$ because the 
gravitino index is trivial. We can therefore use \wittdef\ to show that 
the product of phases around  the boundary is  
\eqn\evalcurv{
\exp\biggl[2\pi i \int_{\Delta \times X} {1\over 6} \tilde G^3 - \tilde G I_8 \biggr] 
= \exp \bigl[ - 2\pi i \half \int_X G \wedge \omega_1 \wedge \omega_2 \bigr] 
}
where $\half$ is  the area of $\Delta$. This proves the desired cocycle 
relation \cocyrel.  
 
The following property of the cocycle $\varphi(\c C, \c \chi)$ 
will be useful in what follows. If $\c \chi_b$ is a topologically trivial 
character then 
%
\eqn\smmirc{
\varphi(\c C, \c \chi+ \c \chi_b) = \varphi(\c C, \c \chi) e^{
-2\pi i \int_X b (\half G^2 - I_8) } 
}
This can be proved 
using the variational formula  \variationalp,  or by using a cobordism argument, 
since for a topologically trivial $\c \chi_b$ we may fill in $S^1_-$ with 
a disk and continue the character as $\c \chi_{rb}$ where $r$ is the radius 
on the disk.  

\subsec{A physical motivation for formula \mphasecf\ } 

The choice of cocycles defining the group lift \lifted\  is 
not unique. Therefore we would like to justify the choice 
\mphasecf. This definition 
of the group lift is necessary if we want to represent 
path integrals in certain twisted topological sectors as traces 
over Hilbert space.

Suppose   $\Psi(\c C_1, \c C_2)$ is the 
path integral on the cylinder $X \times [0,1]$. Then, 
it is the kernel of the operator  $  e^{-\beta H}  $ where 
$H$ is the Hamiltonian and $\beta$ is the length 
of the cylinder. By integrating over the diagonal of the  kernel $\Psi(\c C, \c C)$  
we represent a trace on Hilbert space as a path integral 
over $X\times S^1$: 
\eqn\pathtrac{
Z_a(X \times S^1) = {\Tr}_{\CH_a(X)} e^{-\beta H} 
} 
On both the left and right hand side we are working within 
a topological sector defined by $a\in H^4(X,\IZ)$. 
On the right hand side we have the trace over the 
Hilbert space of wavefunctions with support on 
$C$-fields of characteristic class $a$; on 
the   left-hand side   the path integral is the integral 
 over isomorphism classes of $C$-fields 
with $\c C \in \E_{P(a)}(X\times S^1)$, where $P(a)$ is 
simply pulled back from $X$ to $X \times S^1$.

We would now like to twist this construction using the group 
$\CG(X)$. That is, we would like a path integral 
representation of 
\eqn\twistrce{
{\Tr}_{\CH_a(X)} \CU(g) e^{-\beta H} 
}
where $\CU(g)$ is the unitary operator implementing 
the action of $g$ on the Hilbert space. 
Therefore,  we would like to glue together the 
ends of a cylinder with boundary conditions related by 
an element $g\in \CG(X)$. Thus we consider 
$\Psi(g\c C, \c C)$ for $\c C \in E_{P(a)}(X)$. 
This is an element of a line $\CL_{\c C}^* \otimes \CL_{g\c C}$. 
Now, to give a lift of the group action to $\CL$ is to give 
a specific isomorphism 
\eqn\isomrph{
\CL_{\c C}^* \otimes \CL_{g\c C} \cong \IC
}
Such an isomorphism is necessary if we are to integrate 
over the boundary values $\c C$ to produce a path integral 
over $C$-fields on $X \times S^1$ in the 
topological sector $a + [dt]   a(\c \chi)$. The path 
integral is, after all, just a complex number.  If we 
consider simple field configurations, such as those 
with isomorphism class $[\c C_X] + \c t\cdot \c \chi$ on 
the cylinder then there is no phase factor in the 
evolution on the cylinder, since the evolution is through 
a path of gauge transformations.  On the other hand, the
phase factor in the path integral is \mphasecf\ and this 
phase can only enter through the choice of 
isomorphism \isomrph.

Here again the $E_8$ formalism proves to be very useful
since it allows one to make precise the notation of 
a ``loop $L$ in the space of $C$ fields on $R$'' used in  
 \WittenHC. To do this one relates  
 a group transformation  $(\alpha, \c \chi)$ to an 
$E_8$ bundle automorphism \bdlaut\ and uses this to define 
a twisted bundle with connection  on $X \times S^1$
as in \twisbdl\curlya.

\newsec{The definition of $C$-field electric charge }

In electromagnetism the electric current satisfies 
\eqn\eleccur{
d *F = j_e .
}
If $j_e$ has compact support (or falls off sufficiently rapidly) one 
 then defines  $[j_e] \in H^{n-1}_{DR,cpt}(X)$ as  the electric charge of 
the source. On a closed manifold $X$, this must be zero, since $*F$ 
provides an explicit trivialization.

In $M$-theory, the Chern-Simons term $\Phi(\c C)$ is {\it cubic} 
and hence the theory is nonlinear. The equation of motion 
(in Lorentzian signature) is: 
\eqn\meom{
{1\over   \l^3}  d*G = \half G^2  - I_8(g)
}
Thus, background metrics and $C$-fields induce electric charge. As 
an element of DeRham cohomology 
\eqn\melecchrg{
[\half G^2 - I_8] \in H^8_{DR}(X) 
}
is the induced charge. Nevertheless, \melecchrg\ is not a suitable expression for the 
electric charge induced by a $C$-field and metric. 
The quantization law \geflq\ on 
the $C$-field implies, via Dirac quantization, that the 
$C$-field electric charge is valued in {\it integral 
cohomology} $H^8(X,\IZ)$. 
\foot{Moreover, recall that the quantization law on $G$ 
follows from the existence 
of elementary membranes. 
Indeed, the elementary membrane has $C$-field electric charge 
given by the isomorphism  $H_2(X,\IZ) \cong H^8(X,\IZ)$. }
Thus, in order to describe the  electric charge   induced by 
the self-interactions of $C$ (and gravity) we need to 
define   a precise {\it integral} lift of \melecchrg.  
We will do this in the next section by considering the micro gauge 
transformations of section 3.3.

\subsec{The action of automorphisms on $\CL$}

In     section 6.2 we have defined a group  action  
$
(0,\c \chi): \CL_{A,c} \to \CL_{A, c+ \omega(\c \chi)}
$. 
When $\c \chi$ is flat, i.e.,  $\c \chi \in H^2(X,U(1))$ then 
$\omega(\c \chi) = 0 $ and hence  
$(0,\c \chi)\cdot (A,c) = (A,c) .$ These are the automorphisms of 
the objects $(A,c)$ in the categorical approach.

If $\Psi \in \CL_{A,c}$ then 
\eqn\autoact{
(0,\c \chi)\cdot \Psi = \varphi(\c C_X, \c \chi)^* ~ \Psi 
}
and  $\varphi(\c C_X, \c \chi)$
can be a nontrivial phase.   If $\Psi$ is a gauge invariant   
wavefunction nonvanishing at $\c C_X$ then 
the  invariance under the automorphisms of 
$\c C_X$ requires $\varphi(\c C_X, \c \chi)=1$ for all flat $\c \chi$. 
We will now regard flat characters as elements 
$\c \chi\in H^2(X,\IR/\IZ)$. 
The cocycle law  \cocyrel\  then implies that 
on flat characters, $\varphi(\c C_X,\cdot) $ is a homomorphism
\eqn\vhmfi{
\varphi(\c C_X, \cdot): H^2(X,\IR/\IZ) \to U(1) 
}
Poincare duality now implies the existence of an integral class, 
$\Theta_X(\c C_X) \in H^8(X,\IZ)$ such that 
\eqn\cocypair{ 
\varphi(\c C_X, \c \chi) = \exp[2\pi i \langle \Theta_X(\c C_X) , \c \chi \rangle ] 
}
Equation \cocypair\   defines, mathematically, an important 
integral   class $\Theta_X(\c C_X)$.  
Note that {\it   if a gauge invariant wavefunction is nonvanishing at $\c C_X$ 
then }
\eqn\thetazero{ 
\Theta_X(\c C_X) = 0 .
}
This is the first nontrivial consequence of the Gauss law. 
  
Since $\Theta_X(\c C_X)$ is an integral class it  only depends on 
the characteristic class $a$ of $\c C_X$ (and the spin structure of 
$X$). Therefore, we will sometimes denote it as $\Theta_X(a)$.

\subsec{$\Theta_X(\c C) $  is  $C$-field electric charge   }

To interpret $\Theta_X(a)$ let us ``insert'' a membrane wrapping 
a cycle 
$\sigma \in Z_2(X)$ in the boundary $X$. Let us consider 
  the gauge invariance of the new wavefunction $\Psi_{\sigma}(\c C)$. 
Relative to the wavefunction without the insertion of $\sigma$ 
the flat   gauge transformations $\c \chi$ with $\c \chi \in H^2(X, \IR/\IZ)$ 
act on the wavefunction with an extra phase
\eqn\xtraph{
\c \chi(\sigma) = 
\exp\bigl( 2\pi i  \langle PD[\sigma], \c \chi \rangle \bigr) 
}
Membranes carry $C$-field electric charge. 
Comparison of \xtraph\ with \cocypair\ shows that we should interpret 
$\Theta_X$ as the $C$-field charge induced by the background metric 
and $C$-field.

An important consistency check on our argument is the following.
Using  \smmirc\   and comparing with the definition \cocypair\ shows that 
\eqn\consich{
[\Theta_X]_{\IR} = [\half G^2 - I_8]_{DR}
}

The above arguments lead to a simple extension of 
the Gauss law: In  the presence of membranes 
wrapping a spatial cycle $\sigma\in Z_2(X)$, a gauge invariant 
wavefunction can have support at $\c C$ only if 
\eqn\extendgl{
\Theta_X(\c C)  + PD([\sigma]) = 0 
}

We have discussed a necessary condition for the existence of a nonvanishing 
gauge invariant wavefunction. The $E_8$ formalism is very useful for giving the 
full statement of the Gauss law, and provides an interesting interpretation of 
the quantization of ``Page charges.'' This will be discussed elsewhere \toappear.

\newsec{Mathematical  Properties of $\Theta_X(\c C)$}

The class $\Theta_X(\c C)$ is a subtle integral cohomology 
 class defined by $E_8$ $\eta$-invariants, 
and not much is known about it. Here we collect a few known
facts. 

As we have explained above, restriction 
  to topologically trivial and flat C-field 
gauge transformations $\c \chi_{b}$, for $db=0$
shows that in DeRham cohomology we have \consich. 
Since $[G]_{\IR} = a_{\IR} - \half \lambda_{\IR}$, 
we learn that \foot{ Note, incidentally, that this implies that  
there is a  natural  integral lift of $30\hat A_8 $ defined on 
spin 10-folds. One easily checks that on $C P^5$,
$30 \hat A_8 = (x_{\IR})^4$, where $x$ generates $H^2(CP^5;\IZ)$, 
so $30$ is the smallest multiple of $\hat A_8$ that has 
an integral lift.} 
\eqn\intrms{ 
[\Theta_X(a)]_{\IR} = \half a_{\IR} (a_{\IR}-\lambda_{\IR}) + 30 \hat A_8
}
(To derive this  note that  $30 \hat A_8 = {1\over 8} \lambda^2 - I_8 = {7 \lambda^2 - p_2 \over 48} $.)

There are three basic facts about the integral class $\Theta_X(a)$. First, 
$\langle \Theta_X(a), \c \chi\rangle$ is a spin cobordism invariant
of $(X, a, \alpha)$, where $a\in H^4(X)$ and $\alpha \in H^2(X,\IR/\IZ)$. 
Second, from this and a computation of spin bordism groups 
one can prove that   $\Theta_X$ is a quadratic refinement
of the cup product:  
\eqn\bilinid{ 
\Theta_X(a_1 + a_2) + \Theta_X(0) = \Theta_X(a_1) + \Theta_X(a_2) + a_1   a_2 
}
Third,  $\Theta$ is  ``parity invariant,'' i.e. it satisfies the identity 
\eqn\parityid{
\Theta_X(a) = \Theta_X(\lambda-a).
}

In order to prove \parityid\ we   define a {\it parity transformation} 
on $\c H^4_{\half \lambda}$ via 
\eqn\paritytmn{
 \c \chi^\CP (\Sigma) =  \left( \c \chi(\Sigma)  \right)^*
}
The parity transform takes 
\eqn\partyag{
\eqalign{
& a \to \lambda-a\cr
& G \to - G. \cr}
}
Under this transformation the M-theory phase transforms as 
\eqn\prtytmn{
\Phi_W([\c C]^{\CP};Y)  = (\Phi_W([\c C];Y)^*
}
This follows since, if $[\c C_Z]$ extends $[\c C]$ then 
$[\c C_Z]^{\CP}$ extends $[\c C]^{\CP}$, and 
$G([\c C_Z]^{\CP}) = - G([\c C_Z])$. Now we simply note 
that the integral in \wittdef\ is odd in $G_Z$. 
Applying this observation to the case $Y= X \times S^1$ we 
see that   
\eqn\paritywitt{
\Phi_{\rm W}( [\c C_X]^\CP - \c t\cdot \c \chi; X \times S^1_-) 
= 
\left( \Phi_{\rm W}( [\c C_X] + \c t\cdot \c \chi; X \times S^1_-) \right)^*
}
where we have made a parity transformation $t \to - t$ on the circle. 
Now, \parityid\ follows from \paritywitt.

The cobordism invariance of $\langle \Theta_X(a), \c \chi\rangle$ and the 
bilinear identity \bilinid\ allows us to   compute $\Theta_X(a)$ in 
several simple examples. 
 To choose but one example 
take   $X = L(3, p_1) \times L(3,p_2)  \times S^4$, where 
$L(3,p)$ is any three-dimensional Lens space $S^3/\IZ_p$. 
Let $v$ be a generator of $H^4(S^4,\IZ)$.
If $b_i \in H^2(L(3,p_i))= \IZ_{p_i}$ (we omit pullback symbols)  then 
\eqn\xplii{
\tilde \Theta(v+ b_1 b_2) =\tilde \Theta(v) + \tilde \Theta(b_1 b_2) + v  b_1   b_2 
}
where $\tilde \Theta_X(a):= \Theta_X(a)- \Theta_X(0)$.
Now $\tilde \Theta(b_1b_2)=0$ because we can fill in $S^4 = \p B_5$ and extend 
the Cheeger-Simons character over this manifold. At the same time we can 
extend the class $\alpha \in H^2(X, U(1))$. Similarly, $\tilde \Theta(v)=0$. 
Here we fill in one of the two Lens spaces  (any oriented $3$-manifold has 
a bounding oriented $4$-manifold). So,  we conclude: 
\eqn\xpliii{
\tilde \Theta(v+ b_1 b_2) =v  b_1   b_2 
}
where we used that the spin bordism group is trivial in 3 dimensions. 
 Equation \xpliii\  is potentially relevant to 
$G_2$ compactifications and 5-brane physics.
 
It would be helpful for topological investigations of $M$-theory if further methods were 
developed to compute the subtle integral class $\Theta_X(a)$.

\newsec{$\Phi$ as a cubic refinement, with applications to integration over flat $C$-fields }

\subsec{The sum over flat $C$-fields as a subintegration in the path integral for $\Psi(\c C_X)$. } 

In this section we make some comments on the nature of the 
wavefunction $\Psi(\c C_X)$ associated to a manifold 
$X$ with boundary $Y$. 

The sum over the gauge copies of $\c C_Y$ in the path 
integral automatically projects onto gauge invariant 
wavefunctions. Therefore, the wavefunction will vanish 
unless $\Theta_X(\c C_X)=0$. We restrict attention to 
these components. 

The wavefunction will be a sum over topological sectors. 
These are labelled by extensions $\tilde a$ of the 
characteristic class of $\c C_X$. Let $\iota: X\hookrightarrow Y$, 
then we sum over $\ker \iota^* \subset  H^4(Y,\IZ)$. By the 
long exact sequence this is equivalent to a   sum over 
\eqn\relatv{
H^4(Y,X;\IZ)/\delta H^3(X;\IZ)
}
where $\delta$ is the connecting homomorphism. 

For each extension $\tilde a$ of $a\in H^4(X;\IZ)$ we 
choose an extending bundle $P_Y$ of characteristic  class 
$\tilde a$. We can fix the $\Omega^1(\adp)$ gauge 
symmetry by choosing an extension $(\tilde A_0, \tilde c_0)$ 
of $(A_X, c_X)$. All other extensions are given by 
$(\tilde A_0, \tilde c_0 + c_Y)$ with 
$c_Y\in \ker \iota^*\subset \Omega^3(Y)$. Let us 
denote this space by $\Omega^3(Y,X)$. 
The integral \frmlwve\ reduces to an integral over 
$\Omega^3(Y,X)/\Omega^3(Y,X)_{\IZ}$.

%
%
%

In the path integral   we  integrate over the 
compact space of harmonic forms
\eqn\compsct{
\CH^3(Y,X)/\CH^3(Y,X)_{\IZ}
}
where $\CH^3(Y,X):= \ker \iota^*$ restricted to $\CH^3(Y)$, and 
the addition of such fields yields no cost in action. That is, 
the real part of the Euclidean action is unchanged by the 
addition of such fields.

In the  sum over \relatv\ one can (noncanonically) split the 
integration over the torsion subgroup and a lattice. The
sum over the torsion subgroup can be combined with the integral 
over \compsct\ to produce an integral over $\ker \iota^*$ 
applied to $H^3(Y,U(1))$. 
Using the   topological considerations of this paper
one can make  some exact 
statements about the nature of this subintegration. In the 
next section we spell out these statements.

\subsec{The cubic refinement law}

In order to study the sum over flat $C$-fields we will need a 
result which is interesting and important in its own right,\foot{%
Indeed, this is the starting point for \freedhopkins.} 
namely the interpretation of the $M$-theory phase as a cubic 
refinement of a trilinear form on $\c H^4(Y)$. 

The product of differential characters on $Y$ together with 
evaluation on $Y$ leads to a 
trilinear form $( \c a_1 \c a_2 \c a_3)(Y)\in U(1)$. 
\foot{The product of differential characters  is described in 
\HopkinsRD. }
The  cubic refinement law states that 
\eqn\cubicrefine{
{
\Phi([\c C] + \c a_1 + \c a_2 + \c a_3;Y) \Phi([\c C] + \c a_1 ;Y ) 
\Phi([\c C]  + \c a_2;Y  ) \Phi([\c C]   + \c a_3;Y)
\over 
\Phi([\c C] + \c a_1 + \c a_2 ;Y ) \Phi([\c C] + \c a_1  + \c a_3;Y) 
\Phi([\c C]  + \c a_2 + \c a_3;Y) \Phi([\c C] ;Y ) 
}
=( \c a_1 \c a_2 \c a_3)(Y) 
}
Recall that $\Phi$ descends to shifted differential 
characters, so it makes sense to take  
$\c a_i \in \c H^4(Y)$, while  $[\c C]$ is a shifted 
character. 
In order to prove \cubicrefine\ note that if 
 $\c C, \c a_i$ all extend 
simultaneously on the same spin 12-fold then we 
may use  Witten's definition \wittdef\ as an integral 
in 12 dimensions. The cubic refinement law follows from 
the simple algebraic identity 
\eqn\cubicform{
\eqalign{
&
{1\over 6}(a + x + y + z)^3 - 
{1\over 6}(a + x + y)^3 - {1\over 6}(a+ y + z)^3 - {1\over 6}(a + x + z)^3 \cr
& + 
{1\over 6}(a+x)^3 + {1\over 6}(a+ y)^3 + {1\over 6}(a + z)^3 - {1\over 6} a^3 
= xyz \cr}
}
It thus follows that if we can simultaneously extend the 
differential characters  $[\c C], $ and $\c a_i$ then we have 
\cubicrefine. It follows from the $E_8$ model for the $C$-field 
that, if we can extend the characteristic classes of the 
characters then we can extend the entire character. Therefore 
we consider the purely topological problem of extending 
the class $a(\c C)$ together with the classes $a_i$. 
The extensions 
of the individual classes exist by Stong's theorem. 
Next, consider the obstruction to the existence of an extension 
of an 11-manifold together with a pair of classes $(Y,a_1,a_2)$. 
The obstruction to finding an extension of 
\eqn\pairsex{
(Y, a_1,a_2 ) + (Y,0,0) - (Y, a_1,0) - (Y, 0, a_2) 
}
is measured by the group   $\Omega^{spin}_{11}(K(\Z,4)\wedge 
K(\Z,4))$. Similarly,  the obstruction for  triples $(Y,a_1,a_2,a_3) $ 
lies in  $\Omega^{spin}_{11}(K(\Z,4)\wedge K(\Z,4)\wedge K(\Z,4))$
and that for quartets lies in $\Omega^{spin}_{11}(\wedge^4 K(\Z,4))$.
All  of these groups can be shown 
to vanish by a simple application of the Atiyah-Hirzebruch spectral 
sequence.  

\subsec{Summing over the flat $C$-fields}

We can now apply the cubic refinement law to learn some 
facts about the sum over flat $C$-fields on $Y$. 
For simplicity we will consider the case of $\p Y = \emptyset$.

As we discussed, evaluating the sum over flat $C$-fields reduces 
to evaluation of the 
integral 
\eqn\sumflati{
\int_{H^3(Y,U(1))} [d\c C_f] \Phi(\c C + \c C_f;Y) 
}
where $[d \c C_f]$ is the natural measure given by the 
Riemannian metric. Recall that $H^3(Y,U(1))$ is 
a  disjoint union of connected tori. The connected 
component of the identity is $H^3(Y,\IZ)\otimes U(1)$, 
and the quotient by this subgroup is isomorphic 
to the torsion group $H^4_{T}(Y)$. 
\foot{We will denote the torsion subgroup of $H^p(Y,\IZ)$ 
by $H^p_T(Y)$. } 
To evaluate this we first 
integrate over the subgroup 
of topologically trivial flat $C$-fields, writing the sum \sumflati\ as:  
\eqn\sumflatii{
\sum_{H^4_T(Y)}
\int_{H^3(Y)\otimes U(1)} [d\c C_f] \Phi(\c C + \c C_f;Y) 
}
Using the   variational formula for the 
topologically trivial flat fields we learn that  the 
integral over the topologically trivial 
flat characters $H^3(Y)\otimes U(1)$ vanishes unless  
\eqn\eomwhy{
[\half \tilde G^2 - I_8]_{DR} =0 \in H^8(Y;\IR)
}
Thus, the sum over 
flat $C$-fields projects onto fields allowing a solution 
to the equation of motion. 

When \eomwhy\  is 
satisfied   the function of flat characters $\c C_f \rightarrow \Phi(\c C + \c C_f;Y) $
descends to an interesting  $U(1)$-valued function   
\eqn\remain{
\bar\phi_{[\c C]}(a_T ) :=   \Phi(\c C + \c C_f) 
}
of torsion classes  $a_T\in H^4_T(Y)$.  

The way $\Phi(\c C + \c C_f;Y)$ depends  on $a_T$ is 
not obvious since changing $a_T$ changes the isomorphism class of the 
$E_8$ bundle. It is here that the cubic refinement law 
\cubicrefine\ becomes 
quite useful, since it follows that 
the function $\bar\phi_{[\c C]}$ is a cubic refinement of 
the $U(1)$-valued trilinear form 
\eqn\trilinear{
\exp[2\pi i \langle a_1   a_2 , a_3 \rangle ] 
}
on $H^4_T(Y)$.
Here we have used the torsion pairing $H^8_T(Y)\times H^4_T(Y) \to \IQ/\IZ$. 
In conclusion the sum over flat fields projects onto fields  satisfying 
\eomwhy\ and for such fields we have a further projection onto 
topological sectors such that 
\eqn\torsum{
%
%
\sum_{a\in H^4_T(Y)} \bar\phi_{[\c C]}(a )
}
is nonzero. The sum in \torsum\ is a sum of exponentiated cubic 
forms on a finite abelian group, and is hence  a kind of ``Airy function''
 generalization of   Gauss 
sums.

If we specialize further to the case  $Y = X \times S^1_+$ with 
Ramond, or nonbounding, spin structure on $S^1$ we can go further, 
and make contact with the results of \DiaconescuWY\DiaconescuWZ. 
In this case $H^4_T(Y) \cong H^3_T(X) \oplus H^4_T(X)$ and hence 
the  sum over torsion classes may be arranged as a sum 
\eqn\sumtors{
\sum_{h\in H^3_T(X)} \sum_{a\in H^4_T(X)} \bar\phi_{[\c C]}(h  dt + a)
}
and it is fruitful to study the sum over $H^4_T(X)$ for fixed values of $h$. 

Let us assume that $[\c C]$ is also pulled back from $X$. Then, 
as noted in \DiaconescuWY\ we have a dramatic simplification. 
Let $\c a\in \c H^4(X)$ be a differential character with 
characteristic class $a$ (not necessarily torsion).  Then  
\eqn\dmwres{
{\Phi(\c C + \c a;Y) \over 
\Phi(\c C;Y) } = e^{i \pi [f(a(\c C)+ a) - f(a(\c C))] } 
}
where $f(a)$ is the mod 2 index of the Dirac operator $\Dsl_{V(a)}$ on $X$. 
Moreover, $f(a)$ is a quadratic refinement of a bilinear form: 
\eqn\qrefbf{
f(a_1+a_2) = f(a_1) + f(a_2) +  \int_X r_2(a_1) Sq^2 r_2(a_2) . 
}
where $r_2$ denote reduction modulo two. 
It now follows that  $\bar\phi_{[\c C]}(h  dt + a)$ 
defines a {\it quadratic refinement} as 
a function of $a \in H^4_T(X)$. Indeed,    the cubic refinement 
law reduces to the quadratic refinement law:  
\eqn\newbilfrm{
{
\bar\phi_{[\c C]}(h  dt + a_1+ a_2) \bar\phi_{[\c C]}(h  dt) \over 
\bar\phi_{[\c C]}(h  dt + a_1) \bar\phi_{[\c C]}(h  dt +  a_2) } = 
\exp\left[2\pi i     \langle a_1, (Sq^3 + h ) a_2 \rangle\right] 
}
where we have used the bilinear identity for the $E_8$ mod 2 index 
proved in \DiaconescuWY, and again have used the torsion pairing 
$H^4_T(X) \times H^7_T(X)\to \IQ/\IZ$. 

Thus, for fixed values of $h$ we are averaging a quadratic form over a finite 
abelian group in 
\sumtors.

We now need a little lemma from group theory. 
Let $A,B$ be finite abelian groups, with a perfect pairing 
\eqn\perpar{
A \times B \to \IQ/\IZ
}
which we shall write as $\langle a,b\rangle $.  
Suppose we have a homomorphism $\varphi: A \to B$ such that 
\eqn\queform{
Q(a_1, a_2) = \langle a_1, \varphi(a_2)\rangle
}
is a symmetric bilinear form. Let $f_Q: A \to \IQ/\IZ$ be a quadratic 
refinement of this bilinear form, that is:  
\eqn\quadref{
f_Q(a_1 + a_2)=  f_Q(a_1) + f_Q(a_2) + \langle a_1, \varphi(a_2)\rangle  
}
Note first that, when restricted to $\ker \varphi$, 
$f_Q$ is a character, and hence there is $  P \in B/\im\varphi$ such that the restriction 
of $f_Q$ to $\ker \varphi$ is given by 
\eqn\pdef{
f_Q(a) = \langle a, P\rangle  \qquad a\in \ker\varphi
}
Moreover, $P\not=0$ implies $f_Q(a)\not=0$ for some $a$. 
Now, we have the key statement: {\it The sum  
\eqn\sumqr{
S(f_Q) := \sum_{a\in A} e^{2\pi i f_Q(a)} 
}
is nonzero if and only if $P=0$.}

We now apply the 
above discussion with $A= H^4_T(X), B= H^7_T(X)$ and $\varphi = Sq^3+h$
and 
\eqn\defeff{
{\Phi(\c C + h dt + a )\over \Phi(\c C + h dt)}
= e^{2\pi i f_{\c C,h}(a) }
}
where $f_{\c C,h}(a)$ is  a quadratic refinement of the quadratic form 
 $\langle a_1, (Sq^3 +h) a_2\rangle $ on $H^4_T(X)$.  When restricted to 
$a\in \ker (Sq^3+h)$ we have $f_{\c C,h}(a)= \langle a, P_{\c C,h}\rangle$. 
It follows that 
the   path integral vanishes due to the sum over flat fields unless there 
exists some $C$-field $\c C_0$, such that 
\eqn\criterion{
f_{\c C_0,h}(a) = \langle a, P_0 \rangle =0 
}
for all $a \in \ker(Sq^3+h)\vert_{H^4_T} $. Suppose such a $C$-field exists. We may ask what 
other $C$-fields contribute. Applying once more the cubic refinement law 
we find 
\eqn\repse{
f_{\c C,h}(a) = f_{\c C_0,h}(a) + \langle a, (Sq^3+h) (a(\c C)- a(\c C_0))  \rangle 
}
where $a$ is torsion, but $a(\c C)$ and $a(\c C_0)$ need not be torsion. 
Thus we conclude that the topological sectors $a(\c C)$ which contribute to the path 
integral must satisfy:  
\eqn\tew{
(Sq^3 + h)(a(\c C) - a(\c C_0) ) = 0 \qquad \qquad {\rm mod} (Sq^3  + h)H^4_T(X)
}
This simplifies (considerably) the discussion of the 
``Gauss law'' in   \DiaconescuWY\ and generalizes it to arbitrary torsion 
 $h$-fields. 

Using a similar strategy one can easily derive an $SL(2,\IZ)$ ``equation of motion'' 
for the torsion components of the $C$-field on $11$-manifolds of the type 
$Y= X_9 \times T^2$ where $T^2$ carries the RR (a.k.a. odd, or nonbounding) 
spin structure. (The reader should compare our discussion with 
sec. 11 of \DiaconescuWY.) Choose coordinates $t_1,t_2$ on $T^2$ with 
$t_i \sim t_i +1$ and consider $C$-fields of the type $\c C = \c C_0 + 
\c g\cdot \c t_1 + \c h \cdot \c t_2$ where $\c C_0, \c g, \c h$ are all 
pulled back from $X_9$ and, for clarity, we will drop all pullback symbols. 
The fields $\c g, \c h$  can be interpreted in type IIB string theory 
with $\omega(\c g) = G_{r} $ the fieldstrength of the RR 3-form and 
$\omega(\c h) = H_{ns}$ the NSNS 3=form fieldstrength. 

The sum over the topologically trivial flat $C$-fields imposes \eomwhy, 
which in turn shows that 
\eqn\toptrivcss{
[G_0\wedge H_{ns}]_{DR} = [G_0 \wedge G_{r}]_{DR} = [G_r\wedge H_{ns}]_{DR}=0
}
where $G_0  = \omega(\c C_0)$. The equations \toptrivcss\ are indeed standard 
consequences of the supergravity equations of motion. However, once these 
equations are satisfied there is a further constraint on the characteristic 
classes $a_g := a(\c g)$ and $a_h = a(\c h)$. Note that these classes need not 
be torsion classes, although, by \toptrivcss\ $a_g   a_h$ is a torsion class. 
Combining the cubic refinement law \cubicrefine\ with the bilinear identity 
\qrefbf\ we find 
\eqn\sltzee{
{\Phi(\c C_0 + 
\c g\cdot \c t_1 + \c h \cdot \c t_2+ \c C_f; Y) 
\over 
\Phi(\c C_0 + 
\c g\cdot \c t_1 + \c h \cdot \c t_2; Y) 
} = \langle a_f , Sq^3(a_g + a_h) + a_g   a_h \rangle e^{i\pi f(a_f) } 
}
where $\c C_f $ is a flat character on $X_9$, $a_f = a(\c C_f)$ and 
$f(a_f)$ is the mod $2$ index on $X_9 \times S^1_+$. By the bilinear identity 
$e^{i \pi f(a_f)} = \langle a_f,P\rangle$ where $P$ depends only on the 
topology and spin structure of $X_9$. In particular, it is independent of 
$\c C_0, \c g, \c h$. Thus, we arrive at the $SL(2,\IZ)$ invariant 
equation of motion: 
\eqn\sltzeeii{
Sq^3(a_g) + Sq^3(a_h) + a_g   a_h = P 
}
While this does not fully resolve the puzzle in section 11 of 
\DiaconescuWY, it does constitute progress.

\newsec{Application 1: The 5-brane partition function }

In addition to the electrically charged membranes, $M$-theory 
has magnetically charged 5-branes. 
In \WittenHC\WittenVG\ Witten has analyzed in detail 
topological considerations concerning the 5-brane partition 
function. In particular, he stated topological 
conditions necessary for the construction of a nonzero partition 
function. In this section we make contact with his work and 
the related work of Hopkins and Singer \HopkinsRD. In particular
we interpret a certain anomaly cancellation condition of 
Witten's  in terms of the class $\Theta_X(a)$.

As a preliminary, let us discuss some geometrical facts. The worldvolume of the 
5-brane is denoted by   $W$. We assume $W$ is 
compact and oriented and 
 embedded in an 11-dimensional 
spacetime $\iota: W \hookrightarrow Y$.  We  may identify a tubular neighborhood of $W$ in 
$Y$  with the total space of the 
normal bundle $N\to W$. The unit sphere bundle of radius $r$,  $X=S_r(N)$ is 
then an associated $S^4$ bundle $\pi: X\to W$, and we may 
construct an $11$-manifold $Y_r$ with boundary $X$ by removing the 
  disk bundle of radius $r$,  $Y_r = Y- D_r(N)$. 

If $X$ is oriented, compact, and spin, while $W$ is orientable and compact, 
then one can show that the Euler class  of the 
normal bundle  vanishes and hence 
 the Gysin sequence simplifies to give a set of short exact 
sequences 
\eqn\seses{
0 ~ \rightarrow ~ H^k(W,\Z) 
~ {\buildrel \pi^* \over \rightarrow } ~ 
H^k(X,\Z) 
~ {\buildrel \pi_* \over \rightarrow } ~ 
H^{k-4}(W,\Z) ~ \rightarrow 0 .
}
From this we conclude that $\pi^*: H^3(W,\Z)\to  H^3(X,\Z)$
is an isomorphism. Moreover, 
\eqn\fourclss{
H^4(X,\Z) \cong H^4(W,\Z) \oplus \Z 
}
The isomorphism is noncanonical, indeed,  
 a choice of splitting is  defined by a choice of 
global angular form $\upsilon \in H^4_{cpt}(N)$ with 
$\pi_*(\upsilon)=1$. The general degree four class on 
$X$ can then be written 
\eqn\genfrcl{
a = \pi^*(\bar a) + k \upsilon.
}
with $k\in \Z$. Finally, it follows from the above that 
\eqn\isotori{
H^3(X,U(1)) \cong H^3(W, U(1)).
}

\subsec{Statement of Anomaly inflow}

The 5-brane is a magnetic source for the $C$-field. This means that if $S^4$ is a small 
sphere linking $W$ in $Y$ then 
\eqn\finve{
\int_{S^4} G = k 
}
for a ``5-brane of charge $k$.'' Here $k$ is a nonzero integer. 
In the $E_8$ model for the $C$-field this means that there is an 
$E_8$ instanton on the 4-sphere of instanton number $k$. 
Consequently, the bundle $P(a)\to Y_r$ 
cannot be smoothly prolonged to a bundle over $Y$. There are two ways to approach 
this difficulty. First, one may note that it is natural to associate 
the magnetic current in differential cohomology $\c \delta(W) \in \c Z^5(Y)$   
to a single 5-brane wrapping $W$ and view the $C$-field as a $4$-cochain giving rise 
to this class.   A second point of view, which we shall adopt here, 
 proceeds by dividing up the physical
system 
into two subsystems, the brane and the exterior. More precisely, 
we divide up $Y$ into a small tubular neighborhood $D_r(W)$ of $W$ 
and the complement $Y_r$ of this neighborhood. The two regions 
overlap on the 4-sphere bundle $X$.  
 The exterior of the 5-brane, $Y_r$ is referred to as the 
``bulk'' and is described by 11-dimensional supergravity. 
In particular, the $C$-field path integral  over $Y_r$ 
is the wavefunction 
\eqn\extern{
\Psi_{\rm bulk} \in \Gamma(\CL \to Met(X) \times  \E_P(X) ) 
}
discussed throughout this paper. 

In order to define the ``partition function of $M$-theory 
with a 5-brane wrapping $W$'' we must define the 5-brane 
partition function $Z_{M5}$ together with a well-defined 
pairing 
\eqn\pairing{
\langle Z_{M5}, \Psi_{\rm bulk} \rangle
}
which can 
be integrated over, for example, the space of $C$-fields on $X$ 
to produce  the full partition function. 
The existence of a well-defined,  {\it gauge invariant function}  \pairing, 
is the so-called anomaly inflow cancellation 
statement.   Note, in particular, that $Z_{M5}$ must be a section of 
a dual line bundle to $\CL$. 

Note that, since the 5-brane is a brane, 
$Z_{M5}$ should only depend on local data near $W$.  
In terms of the metric, it should 
be a function of the induced metric on $W$ and its first 
few normal derivatives, such as the 
induced connection on the normal bundle $NW$.  Moreover, 
$Z_{M5}$ should only depend on the ``$C$-fields on 
$W$.''  For this reason we should consider the 
limit that the radius $r$ of the $4$-sphere bundle 
goes to zero in discussing \pairing. 

The gauge equivalence classes of  ``$C$-fields on $W$'' 
will be shifted differential characters 
$[\bar C] \in \c H^4_{\half \lambda_W} (W)$. In order to make the anomaly 
inflow statement precise we must relate $C$-fields on $W$ 
to $C$-fields on $X$. To that end we  {\it choose} a $C$-field 
$\c C_0$ so that we can map 
\eqn\ceewm{
i: \c H^4_{\half \lambda_W} (W) \to \c H^4_{\half \lambda_X} (X)
}
via $i[\bar C] = [\c C_0] + \pi^*[\bar C] $. The choice of 
basepoint $\c C_0$ is not canonical. Note that it should be 
an {\it unshifted} character since $\pi^*(\lambda_W) = \lambda_X$.

The map \ceewm\ is compatible with the 
isomorphism  
\eqn\isotori{
J^3(W) \to J^3(X)
}
where $J^3(X)$ is the connected component of the identity in 
$H^3(X,U(1))$. 
Therefore, the construction of the line $\CL$ over $J^3(X)$ 
can be used to construct a line $\CL^{-1}$ over $J^3(W)$. 

In section 3.2 of \WittenHC\ (see especially p. 125) Witten uses 
the anomaly-inflow statement to determine the line of which $Z_{M5}$ 
should be a section. 
He then notes that Hodge $*$ defines a complex structure on 
$J^3(W)$ such that the curvature form of $\CL^{-1}$
\eqn\mlncrvi{
\omega(x_1, x_2) := -2\pi \int_X      x_1   x_2
}
is of type $(1,1)$. In this 
complex structure $\CL^{-1}$ can be given the structure of 
a holomorphic line bundle  
which in turn gives  
$J^3(W)$ the structure of  a principally polarized abelian variety. 
The partition function $Z_{M5}$ is then identified, up to a metric-dependent constant, 
with the unique holomorphic section of this line bundle. The condition 
of holomorphy is interpreted as the condition of self-duality of the 
3-form fieldstrength of the 5-brane.

\subsec{Intrinsic and extrinsic definitions of the 5-brane partition function}

In \WittenHC\WittenVG, 
Witten has in fact given {\it two} definitions of the 5-brane partition 
function, which we will refer to as the ``intrinsic'' and ``extrinsic'' 
definitions. The extrinsic definition is the one based on anomaly 
inflow, as reviewed in the previous section. The intrinsic definition 
proceeds only from the data of a compact oriented  Riemannian  $6$-manifold $W$.  

In the intrinsic definition we consider $J^3(W)$ as a complex manifold 
with complex structure induced by $*$. We seek to define a holomorphic 
line bundle $\CL_{M5} \to J^3(W)$ with Hermitian metric and 
curvature  given by \mlncrvi. In order to define the holomorphic line one 
must find  a function 
\eqn\wittncocy{
\Omega_W: H^3(W,\IZ) \to \IZ_2
}
which satisfies the cocycle relation: 
\eqn\wittncocyi{
\Omega_W(x_1+x_2) = \Omega_W(x_1) \Omega_W(x_2) \exp[i \pi \int_W x_1   x_2] 
}
Different choices of $\Omega_W$ correspond to lines $\CL_{M5}$ differing by tensoring with 
a flat holomorphic bundle of order 2. 

In \WittenHC\WittenVG\ Witten constructs such a function $\Omega_W$,  when 
$W$ is $Spin^c$, making use of the $Spin^c$ structure. 
\foot{This construction is given in section 5.2 of \WittenVG: One extends 
$W\times S^1$ to a $Spin^c$ $8$-fold $\p B = W\times S^1$, and 
simultaneously extends $x   [dt] $ to a class $z$. Then 
$\Omega_W(x):=\exp[i \pi \int_{B_8} (z^2 + \lambda  z)]$, where 
$2\lambda = p_1 - \alpha^2$ and $r_2(\alpha) = w_2(B_8)$. }
It follows from \wittncocyi\ 
 that $\Omega_W$ is a homomorphism $H^3_T(W,\IZ) \to \IZ_2\hookrightarrow \IQ/\IZ$ and hence 
by Poincar\'e duality $\Omega_W(x) = \langle \theta, x\rangle$, for $x\in H^3_T(W)$,
is pairing with a distinguished element $\theta\in H^4_T(W)$. In 
\WittenVG,  eqs. 5.5 et. seq.,
 Witten then points out that in constructing a theta function as a sum 
over $H^3(W,\IZ)$, if $\theta\not=0$, the sum will vanish, indicating the 
presence of an anomaly. Thus, Witten's anomaly cancellation condition is $\theta=0$. 
In the next section we will interpret the class $\theta$ in terms of 
$\Theta_X(a)$.

\subsec{The decoupling conditions}

It is widely believed that   5-brane theories exist
independently of $M$-theory as  distinguished $(2,0)$-superconformal 
field theories in six dimensions. The ``simple'' theories
are expected to be classified by ADE gauge groups and reduce, 
upon compactification on a circle and upon taking a long distance 
limit, to 5-dimensional nonabelian gauge theories with 16 supersymmetries. 
Nevertheless, the only known definition of the 
 5-brane theory is that given by taking   a decoupling limit from 
the theory of the 5-brane in 11-dimensional supergravity.  
Therefore, we should only expect to be able to give an 
``intrinsic definition'' to the 5-brane partition function 
when the brane can be consistently decoupled from the bulk. 

We will now interpret Witten's torsion anomaly condition $\theta=0$ as 
a necessary condition for decoupling the 5-brane from the bulk. 
Let us now recall the interpretation of 
$\Theta_X(\c C)$ as $C$-field charge. 
If $\Theta_X(\c C)\not=0$ then we cannot decouple the 
5-brane from the bulk. Effectively, open 2-branes must end on the 5-brane
to satisfy charge conservation, and these spoil decoupling. 
 {\it 
Therefore, physical reasoning implies that 
$\Theta_X(\c C) =0$ is a necessary condition 
for decoupling of the $M5$ brane from the bulk, i.e. 
a necessary condition for the existence of a nonzero  
``5-brane partition function.'' }
 
In order to state the decoupling condition in terms intrinsic to $W$  
let  us consider   the integration 
over the fiber:  $\pi_*(\Theta_X(a)) \in H^4(W_6, \IZ)$. 
We use   \genfrcl\ to conclude that that  $a = \pi^*(\bar a) +  \upsilon$
for a single brane. Then we may split (noncanonically)  $\pi_*(\Theta_X(a))=0 $ into its 
torsion and nontorsion components.  

In DeRham cohomology $\pi_*(\Theta_X(a))=0 $ implies that $\bar a = 0$. 
This is the well-known condition that the ``$C$-field on $W$'' must have a 
fieldstrength which can be written as
\eqn\fidls{
 \bar G = dh 
}
for some globally well-defined $3$-form $h$. The form $h$  is interpreted as the 
fieldstrength of the chiral 2-form field on $W$. 

When \fidls\ is satisfied   $\pi_*(\Theta_X(a))$ is a torsion class. 
As we have seen, it  is precisely this class which obstructs the definition of a well-defined 
line bundle $\CL \to J^3(X)\cong J^3(W)$. When $\Theta_X(a)=0$ we can identify 
\eqn\wittsow{
\Omega_W(x) = \varphi(\c C_0, \c \chi).
} 
Here $\c \chi$ is a character such that $\omega(\c \chi)=x$. Of course, such 
lifts are ambiguous by flat characters, but, precisely because $\Theta_X(a)=0$, 
this ambiguity drops out of the right hand side of \wittsow. This is 
in accord with the argument in \WittenVG, section 5.3 . (The latter argument 
relied on several topological restrictions on the normal bundle of $W$ in $Y$. ) 

In general, the anomaly inflow argument implies that, for $\alpha \in H^2(W,U(1))$ we should 
identify $\langle \a, \theta \rangle = \Omega_W(\beta_*\alpha)$, where $\beta_*$ is 
the Bockstein homomorphism, with $\Phi(X \times S^1_-, \c C + \c t\pi^*\alpha)= 
\langle \pi^*(\alpha),\Theta_X(\c C) \rangle = \langle \alpha, \pi_*\Theta_X(\c C) \rangle$, 
and thus we conclude that $\theta= \pi_*(\Theta_X(\c C))$.

\newsec{Application 2: Relation of $M$-theory to $K$-theory}

In the previous section we have shown that if  
$\pi_*(\Theta_X)\not=0$ then there must be $2$-branes 
ending on a $5$-brane.  In type II string theory there is 
an   analogous  statement for $D$-branes. If a $D$-brane wraps a 
worldvolume $W$ in isolation, i.e. with no other $D$-branes 
ending on it, then it is necessary that: 
\eqn\iiacond{
(Sq^3 + [H])PD(W) =0 .
}
This condition is closely related  to the K-theoretic classification of D-branes, 
as explained in 
\DiaconescuWY\DiaconescuWZ\MaldacenaXJ\evslin\MooreVF.  
If $Y_{11}$ is $S^1$-fibered 
over a 10-manifold $U_{10}$, then   we expect a relation between $M$ theory 
on $Y$ and   type II 
string theory on $U_{10}$. Therefore, in this situation we expect 
a relation between $\Theta_X(a)$ and the left hand side of \iiacond. 
We will now show that this is indeed the case. 
 
For simplicity, suppose  $Y_{11} = U_{10} \times S^1$.
Then $\p Y_{11}= X_{10}$, $\p U_{10}= V_9$, so 
$X_{10}= V_9 \times S^1$. A characteristic class of $\c C_X$ on $X_{10}$ 
can be written as: 
\eqn\updown{ 
a = \pi^*(\bar a) + \pi^*([H])   [dx^{11}] 
}
where $\bar a\in H^4(V_9;\IZ)$,  $[H]\in H^3(V_9;\IZ)$, 
and $\pi: X_{10} \to V_9$ is the projection. 

The first result is that if $X_{10}= V_9 \times S^1$ 
has Ramond  (nonbounding) spin structure on $S^1$ 
and $[H]=0$ then 
\eqn\pixa{
\pi_*(\tilde \Theta_X(a)) = Sq^3(\bar a) 
}
Thus the Gauss law and the $K$-theoretic restriction 
on the RR fluxes are indeed closely related.

To prove \pixa\  we compute  the pairing of the left 
hand side with 
$\alpha \in H^2(V_9, U(1))$. This is   
\eqn\paralphs{
\langle \Theta_X(a), \pi^*(\alpha)\rangle=
\Phi(p^* \check C + \check t  p^*(\pi^*\c \alpha), X \times S^1_-)
}
where $\c C \in \c H^4(X_{10})$, $p: X_{10} \times S^1_- \to X_{10}$, and the 
coordinate on $S^1_-$ is $t$. 

If $\c C$ is pulled back from $V_9$ then we may regard the $11$-manifold 
$Y= X_{10} \times S^1_-$ instead as $Y= X' \times S^1_+$, with $X' = V_9\times S^1_-$. 
Then we can apply to result of  \DiaconescuWY\ to obtain
\eqn\eiant{
\langle \pi_* \tilde\Theta_X(a), \alpha\rangle = e^{i \pi [f(\bar a + dt  \beta(\pi^*\alpha)) 
- f( dt  \beta(\pi^*\alpha)] } 
}
where $\beta:H^2(V_9,U(1)) \to H^3_T(V_9) $ is the Bockstein
and $f(a)$ is the $E_8$ mod 2 index on $X'= V_9 \times S^1_-$. 
Here we have used \DiaconescuWY\ eq. 8.24  and the fact that 
$\beta(\alpha)$ is flat, hence the local term cannot 
contribute because the local density does not contain $dt$. 
Now we use the bilinear identity of \DiaconescuWY. $f(\bar a)=0$ since the 
$S^1_- $ is a bounding spin structure.  Therefore 
\eqn\mtokay{
\langle \pi_*\tilde \Theta_X(a), \alpha\rangle 
= e^{  i \pi \int_{V_9 \times S^1_-} \bar a Sq^2 dt   \beta (\alpha )}
 = \langle Sq^3(\bar a), \alpha \rangle 
}
where in the second equality we used  
 the pairing $H^7(V_9,\IZ) \times H^2(V_9, U(1)) \to U(1)$. 
This establishes \pixa. 

It is also of interest to know $\pi_*(\Theta_X(0))$. 
The style of argument above shows that 
\eqn\thetzer{
\langle \alpha, \pi_*(\Theta_X(0))\rangle = e^{i \pi f(\beta(\alpha)  dt) } 
}
This formula shows that $\pi_*(\Theta_X(0))$ is at most two-torsion.  
There seems to be   no elementary formula for this mod two index. 
If $V_9= X_8 \times S^1_+$ then one can show that
\foot{Let $+$ denote the nonbounding, $-$ the bounding spin structure on $S^1$. 
Note that $f_{--}(b  dt_1)=f_{--}(b   dt_2)=0$ where the subscript 
denotes the spin structure. Note that under a diffeomorphism $(t_1,t_2) \to (t_1, t_1+t_2)$ 
we have $f_{+-}(bdt_2) = f_{--}(bdt_1 + b dt_2)$. Now apply the bilinear identity.} 
  $f(b  dt) = \int r_2(b) Sq^2 r_2(b)$
for $b\in H^3(X_8,\IZ)$. The latter expression is, in general, nonzero (e.g. for 
$X_8 = SU(3)$ and $b$ a generator of $H^3(SU(3);\IZ)$ ), and hence we conclude that 
$\pi_*(\Theta_X(0))$ is in general nonzero. 

%
%

Now let us turn to the inclusion of an $[H]$ flux. 
It is straightforward to show 
\eqn\bdrydmwiii{
\pi_*(\Theta_X(a)_{DR}) =[H]_{DR} \wedge \bar a_{DR}
}
There are two ways to lift this to a statement about the 
integral classes. First, using the bilinear identity 
\bilinid\ (with $a_2 \to a_2-a_1$, together with  
\pixa) we find that
 if $a_i =\pi^* \bar a_i + \pi^* [H]   [dx^{11}]$ then 
\eqn\bdrydmwv{
\pi_*( \Theta_X(a_1)- \Theta_X(a_2)) = (Sq^3 + [H])   (\bar a_1 - \bar a_2) 
}
Moreover, using \parityid\  we also have 
\eqn\bdrydmwiv{
\pi_*(2\Theta_X(a)) =   [H]   ( 2\bar a - \bar \lambda)
}
Now,  \bdrydmwiv\ and \bdrydmwv\ together 
 amount to  the   ``moral'' statement 
\eqn\moralst{
\pi_*(\Theta_X(a)) ``= {\rm '' } (Sq^3 + [H])(\bar a - \half \bar \lambda) 
}
(it is only a moral statement because the division by $2$ is illegal 
in the presence of 2-torsion). Hence,  
 the Gauss law is in harmony with  the $K$-theoretic
classification of RR fluxes.

The above arguments extend the results of \DiaconescuWY\DiaconescuWZ\ to manifolds with 
boundary. To complete the story one should demonstrate that the
$C$-field wavefunction is related to the RR flux wavefunction in the 
expected manner. This is an interesting problem, but we leave it for 
future work. (Some preliminary remarks appear in \MooreGB.)

\newsec{Application 3: Comments on spatial boundaries  }

One of the primary motivations for developing the $E_8$ 
formalism for the $C$-field is the desire to make precise 
the intuition that the $C$-field is related to some 
kind of topological gauge theory  in the bulk of 11 dimensions,
but
which becomes a dynamical theory upon the introduction of 
boundaries. In this section we make some simple  
comments on one realization of that idea.

Suppose now we have a {\it spatial} boundary $\iota: X \hookrightarrow Y$. 
The adjective ``spatial'' means that we will be regarding 
$X$ as a Euclidean spacetime, rather than as a temporal slice. 
We need to choose boundary conditions on the $C$-field. 
One natural choice of boundary condition on   $\c C = (A,c)$ is 
the condition 
\eqn\splebc{
\iota^*(c)=0. 
}
That is, we restrict attention to the $C$-fields: 
\eqn\restrc{
\E_P(Y,X) := \{(A,c)\in \E_P(Y)\vert \iota^*(c)=0 \}
}

Now, to define a physical theory, we need to describe 
the gauge symmetry. 
It is quite standard for boundary conditions to break 
a symmetry group $G$ to a subgroup $H$. It is thus 
natural to consider the subgroup of $\CG(Y)$ of elements 
that preserve \restrc.   In the present 
case,   the group elements $(\alpha, \c \chi)\in \CG(Y)$ 
which preserve the entire set \restrc\  must satisfy 
$\iota^*(\alpha)=\iota^*(\omega(\c \chi))=0$. Such 
group transformations leave ``too many'' physical 
degrees of freedom. In particular, any two connections 
on $\iota^*(P)$ would be considered to be  gauge inequivalent.

It is here that the categorical viewpoint of section 3.4
becomes quite useful. Let ${\hat P} = \iota^*(P)$. 
There is an obvious restriction functor 
$r: \E_P(Y)//\CG(Y) \to \E_{\hat P}(X)//\CG(X)$.  
The simplest 
way to characterize the category defined by the boundary 
condition \splebc\ is that it is the subcategory of 
$\E_P(Y)//\CG(Y) $ which maps under $r$ to  a category with the morphisms 
of the form $(A,0) \to (A^g, 0)$, for   $g\in {\rm Aut {\hat P}}$. 
To say this a little more formally,   we noted 
in section 3.4 above  that
ordinary bundle automorphisms of ${\hat P}$ define a subgroupoid 
of $\E_{\hat P}(X)//\CG(X)$, in spite of the fact that 
 ${\rm Aut}({\hat P})$ is not a subgroup 
of $\CG(X)$. Let us fix a functor $I$ from the standard 
gauge theory groupoid $\CA({\hat P})//{\rm Aut}({\hat P})$ to $\E_{\hat P}(X)$, 
by choosing $I(A) = (A,0)$ on objects. The groupoid 
of $C$-fields and gauge transformations defined by 
\splebc\ is the fiber product $\CE_P(Y,X)$: 
\eqn\fiberprod{
\matrix{
\E_P(Y)//\CG(Y) &  {\buildrel r \over \rightarrow }  &  \E_{\hat P}(X)//\CG(X) \cr
\uparrow &   &  \uparrow I \cr 
\CE_P(Y,X) & \rightarrow &  \CA({\hat P})//{\rm Aut}({\hat P}) \cr}
}

The situation is best described as follows: 
This boundary condition breaks the topological gauge symmetry 
$\CG$. It is not possible to describe it as the breaking of 
a gauge group, but we can view it as a {\it   symmetry 
breaking of groupoids.} Note in particular that while 
$\int_Y  {\tr} F\wedge *F $ is not gauge invariant, and hence 
there are no propagating gauge modes in the interior,  
 $\int_X {\tr} F\wedge *F $ is gauge invariant 
 and therefore, with the above notion of a groupoid of 
fields, we can define a gauge invariant theory with 
dynamical  $E_8$ gauge fields  on the boundary.

\subsec{ Relation to heterotic $M$-theory}

\def\c{\check}

Now that we have explained one way in which 
  dynamical gauge fields on the boundary can be related to the 
topological gauge theory of the $E_8$ gauge field in the interior
we will indicate how it can be compatible with the  
 outstanding example of $M$-theory on a manifold with 
spatial boundary, namely  the   Horava-Witten model \HoravaMA\HoravaQA\ 
of heterotic $M$-theory.

Consider the  $E_8$ model of the $C$-field on an $11$-manifold of the type 
$X \times [0,1]$ and impose boundary conditions 
$\iota^*_L(c) = \iota^*_R(c)=0$. We learned in the previous 
section that  we indeed find dynamical $E_8$ gauge fields on each boundary. However, 
because the 11-dimensional spacetime provides a homotopy of the left and 
right connections the $E_8$ bundles on the boundaries necessarily have 
characteristic classes $a_L = a_R$. Thus, this theory resembles the 
nonsupersymmetric 
model introduced  by Fabinger and Horava \FabingerJD.

The above simple observation raises a challenge to using the 
$E_8$ formalism to describe the Horava-Witten model. 
We may overcome this difficulty as follows. Note that 
$M$-theory should be formulated without use of an 
orientation, but our formulation 
breaks parity automatically since $a \to \lambda-a$ under 
orientation reversal.

We may give a parity invariant formulation of the $M$-theory 
$C$-field by passing from $Y$ 
to $ Y_d  $, the
orientation double cover of $Y$, and defining a $C$-field to be
a parity invariant $E_8$ cocycle on $Y_d$.  Let  $\sigma $ be the
nontrivial deck transformation on $Y_d$, a parity invariant
$E_8$ cocycle is one such that the shifted differential character
satisfies:
\eqn\prtyinvt{
\sigma^*([\c C]) =  [ \c  C]^{\CP}
}
where we recall that the parity transform was defined in 
\paritytmn. 
If  $Y$  is orientable, this amounts to saying that a $C$-field is a
pair  $\{ (A_1, c_1), (A_2,c_2)\} $ such that
\eqn\prtyinvtii{
\c \chi_{(A_1, c_1)} = \c \chi_{(A_2,c_2)}^*      
}

The gauge group is $\CG(Y_d) \times \CG(Y_d)$. This 
gauge invariance preserves \prtyinvtii. 
So, again, no degrees of
freedom are added when $Y$  has no boundary.
 However, on a  manifold with boundary this formulation leads to a natural
boundary condition which again leads to dynamical 
gauge fields on the boundary.   On  $X \times [0,1]$ we
take
\eqn\newbcs{
i^*_L(c_1)=0   \qquad  i^*_R(c_2 ) =0
}
Then, we get a dynamical gauge field  $A_L$  with characteristic class $a_L$
on the left boundary and another one $A_R$ with $a_R$  on the right boundary. 
Note that $\iota^*_L([G])= a_L - \half \lambda$, while 
$\iota^*_R([G]) = - (a_R - \half \lambda)$ and hence   $a_L + a_R = \lambda$, 
as expected.

It is important to note that with our boundary conditions $\iota^*(G) = 
\pm ( \tr F^2 - \half \tr R^2)$, on left- and right- boundaries, 
respectively, while   $d G=0$ throughout the 11-manifold. This is to be contrasted with the 
equation one often finds in 
the literature on heterotic $M$-theory, namely, 
\eqn\otherbian{
d G = \delta(x) \bigl(\tr F(A_L)^2 - \half \tr R^2\bigr) \pm   \delta(\pi-x)\bigl(\tr F(A_R)^2 - \half \tr R^2\bigr).
}
In our view, such a Bianchi identity implies that the boundary 
carries nontrivial magnetic current, and a proper formulation of the $C$-field 
will involve a different geometrical construction from what we have used. 

It is also worth noting that since $\Phi$ is valued in the Pfaffian line bundle 
the   $E_8$ gauge anomalies cancel (locally) in a way which is manifest.  
The gravitational anomalies are more subtle, but we expect that the same will 
be true for them. We hope to discuss this elsewhere \toappear. 
(See \BilalES\ for a different point of view.)

\newsec{ Conclusions and future directions}

In this paper we have given a precise mathematical 
formulation of the $E_8$ model for the $M$-theory 
$C$-field. We have used it to write the Chern-Simons 
term of the $11$-dimensional supergravity action 
and we have used it to describe the Gauss law for the 
$C$-field in precise terms, applicable to topologically 
nontrivial situations. In particular we have shown 
how to define the $C$-field electric charge induced by 
the nonlinear interactions of the $C$ field, and by gravity,
as an integral cohomology class. We have also given 
other applications to a clarification of  the topological conditions 
for the existence of the 5-brane partition function
and to the relation of $M$-theory flux quantization to 
$IIA$ $K$-theoretic flux quantization. Finally, we have 
sketched how one may formulate the topological field theory 
of the $E_8$ gauge field to allow for dynamical gauge fields 
on the boundary of an $11$-manifold. 

Many  related  problems remain open and issues remain to be 
resolved. We will survey some of these problems here.

\bgn
\bul\ {\it Gravitino determinant.} 
There are several nontrivial technical issues related 
to the gravitino path integral. We hope to address these 
elsewhere \toappear.

\bgn
\bul\ {\it Derivation of the $K$-theoretic classification of 
RR fields. } The $K$-theoretic 
formulation of RR fields should be generalized to 
manifolds with boundary and the wavefunction of the
$C$-field in IIA supergravity  compared to the wavefunction of 
$C$-field in $M$ theory. We made some progress in 
section $11$  but the story remains to be completed.

\bgn
\bul\ {\it Is the $E_8$ formalism really necessary ?}
One of the virtues of the $E_8$ formalism is that it 
allows us to define the action of $11$-dimensional 
supergravity, and the Gauss law for wavefunctions 
of $C$-fields on manifolds with boundary. Nevertheless, 
the $E_8$ gauge field plays a purely topological role 
and appears, in some sense, to be a ``fake.'' 
As we have mentioned, the category $\E_P(X)//\CG$ is 
equivalent to the cateogry $\c Z^4_{\half \lambda}(X)$ 
of Hopkins-Singer cocycles. Indeed, there is an 
alternative construction of the $C$-field and 
$M$-theory action which makes no use of $E_8$ 
gauge fields \freedhopkins.  What remains 
to be seen is whether any  of these formalisms is 
really useful for physical investigations, and whether 
they lead to a   useful reformulation of $M$-theory.

\bgn
\bul\ {\it Electric-Magnetic duality}. The $E_8$ formalism 
is seemingly very asymmetric between $5$-branes and $2$-branes.  
Formally, one would expect a dual formulation of the theory in 
terms of Cheeger-Simons characters $[\c C_D] \in \c H^7(X)$. 
 These objects would define the holonomies of the $C$-field on 
5-brane worldvolumes. However, there is 
no obvious $E_8$ model for a dual object $\c C_D$. Since the 
theory is nonlinear, this duality transformation is not 
obvious.

\bgn
\bul\  {\it Parity invariance}. A basic axiom of  $M$ theory is 
that it is   parity invariant. This is what allows us to  gauge 
parity and produce chiral theories such as the heterotic 
string. In this paper, we have assumed that $Y$ is an 
oriented $11$-manifold and we have used the orientation 
heavily in writing integrals of differential forms, and in 
defining  
spinor bundles and $\eta$ invariants. A very interesting 
open problem is the generalization of our formalism 
to unoriented and nonorientable $11$-manifolds.

\bgn 
\bul\ {\it Anomaly cancellation.} In spite of the shortcommings 
noted above, we  believe that 
the present formalism should have many future applications 
to  anomaly cancellation issues connected to 5-branes, 
$G_2$ singularities,  and frozen singularities in $M$-theory \deBoerPX.  
For example, the present formalism leads to a substantial 
simplification of the anomaly cancellation for normal bundle 
anomalies of the M5-brane described in \FreedTG.

\bigskip  
\noindent{\bf Acknowledgements:}   

We would like to thank  J. Harvey, M. Hopkins, D. Morrison, G. Segal, and E. Witten 
 for useful discussions on this material. G.M. and D.F. would like to 
thank H. Miller and D. Ravenel for the invitation to speak at the 
Newton Institute conference on elliptic cohomology.  Portions of this work were 
carried out by G.M. at the Aspen Center for Physics. Portions of this 
work were carried out by   D.F. and G.M. at the Isaac Newton Institute, 
Cambridge and at the KITP, Santa Barbara. 
The work of E.D. and G.M.  is supported in part by DOE grant DE-FG02-96ER40949.  
The work of D.F.  is supported in part by NSF grant
DMS-0305505.

\listrefs

\bye